\pdfoutput = 1
\documentclass[journal]{IEEEtran}
\usepackage{bm}
\usepackage{amsmath}%
\usepackage{verbatim}
\usepackage{multirow,diagbox}
\usepackage{graphicx}
\usepackage{subfigure}
\usepackage{amsfonts}
\usepackage{array}
\usepackage{booktabs}
\usepackage[noend]{algpseudocode}

\usepackage{algorithmicx,algorithm}
\usepackage{epsfig}
\usepackage{algpseudocode}
\usepackage{url}
\usepackage{cite}
\usepackage{color}

\ifCLASSINFOpdf
\else
\fi
\begin{document}
%
\title{Image Segmentation Based on Multiscale Fast Spectral Clustering}
%
%
%

\author{Chongyang Zhang, Guofeng Zhu, Minxin Chen, Hong Chen, Chenjian Wu
\thanks{This work was supported by the National Natural Science Foundation of China under Grant 61801321.}
\thanks{Chongyang Zhang and Chenjian Wu are affiliated with the School of Electronic and Information Engineering, Soochow University, Suzhou, China.}
\thanks{Guofeng Zhu, Minxin Chen and Hong Chen are affiliated with the School of Mathematical Sciences, Soochow Univerity, Suzhou, China.}
\thanks{The corresponding author: Chenjian Wu, E-mail: cjwu@suda.edu.cn}}

\markboth{IEEE TRANSACTIONS ON IMAGE PROCESSING,~Vol.~XX, No.~X, XX~2018}%
{Shell \MakeLowercase{\textit{et al.}}: Bare Demo of IEEEtran.cls for Journals}
\maketitle
\begin{abstract}
In recent years, spectral clustering has become one of the most popular clustering algorithms for image segmentation. However, it has restricted
applicability to large-scale images due  to its high computational complexity.
In this paper, we first propose a novel algorithm called Fast Spectral Clustering  based on quad-tree decomposition. The algorithm focuses on the
spectral clustering at superpixel level and its computational complexity is \bm{$O(n\log n)+O(m)+O(m^\frac{3}{2})$}; its memory cost is \bm{$O(m)$},
where \bm{$n$} and \bm{$m$} are the numbers of pixels and the superpixels of a image. Then we propose Multiscale Fast Spectral Clustering  by improving
Fast Spectral Clustering, which is based on the hierarchical structure of the quad-tree. The computational complexity of Multiscale Fast Spectral
Clustering is \bm{$O(n\log n)$} and its memory cost is \bm{$O(m)$}.  Extensive experiments on real large-scale images demonstrate that Multiscale Fast
Spectral Clustering outperforms Normalized cut  in terms of lower computational complexity and memory cost, with comparable clustering accuracy.
\end{abstract}

%
\IEEEpeerreviewmaketitle

\section{Introduction}
\IEEEPARstart{C}{lustering} is an important method of  data processing with a wide range of application such as topic modeling \cite{Rocco}, image
processing \cite{Zhang}, \cite{Zhou},
medical diagnosis \cite{Yu} and community detection \cite{Rong}. 
and applied to imagesegmentation.
 A variety of clustering algorithms have been developed so far, including prototype-based algorithm \cite{cluster-1}, density-based algorithm
 \cite{cluster-2}, graph theory-based algorithm \cite{cluster-4}, etc. The $k$-means algorithm \cite{kmeans}, a prototype-based algorithm, has the advantage of low
 computational complexity. However, it doesn't work well on non-convex data sets. Density-Based Spatial Clustering of Applications with Noise (DBSCAN) is
 a typical density-based algorithm, but it costs a large amount of memory. Spectral clustering algorithms based on the graph theory are appropriate for
 processing non-convex data sets \cite{Qiang,Filippone} though, it is difficult to be applied to large-scale images due to its high computational
 complexity \cite{Rocco, wang,Tung,He,Semertzidis,Cao}, which is primarily caused by two procedures: 1) construction of the similarity matrix, and 2)
 eigen-decomposition of the Laplacian matrix \cite{Hagen}. The computational complexity of procedure 1) is $O(n^2)$ and that of 2) $O(n^3)$,
 an unbearable burden for the segmentation of large-scale images.\par
In recent years, researchers have proposed various approaches to large-scale image segmentation. The approaches are based on three following strategies:
constructing a sparse similarity matrix, using Nystr\" om approximation and using representative points. The following approaches are based on constructing a sparse
similarity matrix. In 2000, Shi and Malik \cite{shi} constructed the similarity matrix of the image  by using the k-nearest neighbor sparse strategy to
reduce the complexity of constructing the similarity matrix to $O(n)$ and to reduce the complexity of eigen-decomposing the Laplacian matrix to
$O(n^\frac{3}{2})$ by using the Lanczos algorithm. However, its computational complexity and memory cost are still high when their method is applied to
large-scale images. To further reduce the computation time and memory cost, in 2005, T. Cour et al. \cite{T.cour} used multiscale graph decomposition to
construct the similarity matrix. The computational complexity of this algorithm is linear in the number of pixels. Some researchers proposed several approaches based on
Nystr\" om approximation.  In 2004, C. Fowlkes et al. \cite{C.Fowlkes} presented the method based on Nystr\" om approximation, in which only a small
number of random samples were used to extrapolate the complete grouping solution. The complexity of this method is $O(m_1^3)+O(m_1n)$, where $m_1$
represents the number of sample pixels in the image. However, deterministic guarantee on the clustering performance cannot be provided by random sampling
\cite{Qiang}. In 2017, Zhan Qiang and Yu Mao \cite{Qiang} improved the algorithm of spectral clustering based on incremental Nystr\" om by the Nystr\" om
sampling method. Computational complexity was reduced to $O(n^2)+O(Mm_1+nm_1^2)+O(knt)$, where $k$ represents the number of clusters, $t$ represents the
number of the iterations of $k$-means and $M$ is a constant. The following approaches are based on representative points. In 2009, Yan et al. \cite{Yan} proposed the
$k$-means-based approximate spectral clustering method. First, The image is partitioned into some superpixels by $k$-means. Then, the traditional spectral
clustering is applied to the  superpixels. The computation time of the method is $O(k^3)+O(knt)$.  In 2015, Cai et al.  \cite{Deng} proposed a scalable
spectral clustering method called Landmark-based Spectral Clustering (LSC). LSC generates $p$ representative data points as the landmarks and uses the
linear combinations of those landmarks to represent the remaining data points. Its computational complexity scales linearly with the size of problem.\par
In this paper, we first propose a novel  spectral clustering algorithm for large-scale image segmentation based on superpixels called Fast Spectral
Clustering (FSC). Then we enhance the method and present Multiscale Fast Spectral Clustering (MFSC), which is based on the hierarchical structure of the
quad-tree. A brief introduction to MFSC: The superpixels of image $I$ are obtained by quad-tree decomposition during which the hierarchical structure of
the quad-tree is reserved. We propose a ``bottom up" approach: along the hierarchical structure of the quad-tree, we merge child nodes at the fine level
into their parent node at the coarse level by treating the clusters, the segmentation result of child nodes, as the superpixels of the parent node. The
computational complexity of the algorithm is $O(n\log n)$ and its memory cost is $O(m)$.\par
The reminder of the paper is organized as follows. In Section \ref{section:Preliminary}, we introduce the preliminaries to the formulation of our
algorithms from the aspects of Ncut and quad-tree decomposition. In Section \ref{section:Method}, we describe our two algorithms FSC and MFSC and their
respective complexity in detail. Experimental results are shown in Section \ref{section:Experiment}. Finally, we conclude our work in Section
\ref{section:conclusion}.
%
%
%
%

\section{Preliminaries}
\label{section:Preliminary}

\subsection{Normalized Spectral Clustering}
\label{subsection:Pri-sc}
This section gives a brief introduction to K-way Normalized cut (Ncut) proposed by Shi et al. \cite {shi}. Suppose image $I$ contains pixels
$v_1\textrm{,}  \ldots\textrm{,}  v_n$, and the similarity matrix of image $I$ is the matrix $W=(w_{ij})_{n\times n}$, in which $w_{ij}$ denotes the
similarity between pixel $v_i$ and pixel $v_j$ \cite{cluster-3}. According to T. Cour et al. \cite{T.cour}, $w_{ij}$ is defined as follows:
\begin{equation}
\label{equ:wij}
\small
w_{ij}=\left\{\begin{array}{rcl}\sqrt{w_I(i\textrm{,}j)\times w_C(i\textrm{,}j)}+\alpha w_C(i\textrm{,}j)  & & {\| X_i-X_j\|^2 \leq r^2\textrm{,} }\\
0\qquad{}\qquad{}\qquad{} & & {otherwise.}\end{array} \right.
\end{equation}
where
\begin{equation*}
\begin{split}
&w_I(i\textrm{,}j)=e^{-\left. \| X_i-X_j \right \|^2/\sigma_x-\left \| Z_i-Z_j \right. \|^2/\sigma_I}\textrm{,} \\
&w_C(i\textrm{,}j)=e^{\min\limits_{x\in line(i\textrm{,}j)}\left. -\|Edge(x) \right. \|^2/\sigma_C}\textrm{,}
\end{split}
\end{equation*}
where $X_i$ and $Z_i$ denote the location and intensity of pixel $v_i$;  $r$ denotes graph connection radius; $\sigma_x$ and $\sigma_I$ are scaling
parameters; $Edge(x)$ is the edge strength at location $x$;  $line(i\textrm{,}j)$ is the straight line connecting pixels $v_i$ and $v_j$ \cite{T.cour}. If
the straight line connecting the two pixels does not cross the edge of the image, the value of $w_{C}$ will be large, reflecting that the affinity of the
two pixels is high. With the similarity matrix $W$, K-way Ncut clusters the image into $k$ clusters $C=\{C_1\textrm{,} \; C_2\textrm{,}...\textrm{,}C_k\}$ by
solving the following minimization problem \cite{shi}, \cite{tutorial}, \cite{Ng.andrew}:
\begin{equation}
\label{equ:Ncut}
\min\limits_{C}\quad Ncut(C)\textrm{,}
\end{equation}
where
\begin{equation*}
\begin{split}
&Ncut(C)=\frac{1}{2}\sum_{i=1}^k\frac{cut(C_i\textrm{,}\bar{C_i})}{vol(C_i)}\textrm{,}  \\
&vol(C_i)=\sum_{v_i\textrm{,}v_j\in{C_i}}w_{ij}\textrm{,}
\end{split}
\end{equation*}
where $\bar{C_i}=C-C_i$ represents the complement of $C_i$; $cut(C_i\textrm{,}\bar{C_i})=\sum\limits_{v_i\in{C_i}\textrm{,}v_j\notin{C_i}}w_{ij}$ reflects
the connectivity strength between $C_i$ and other clusters; $vol(C_i)$ $(i=1\textrm{,} ...\textrm{,} n)$ is the regularization term preventing the
clustering result from being an isolated pixel.\par
 To solve the above problem,  the matrix $X=(x_{ij})_{n \times k}$ is defined as follows:

\begin{equation}
\label{equ:xij}
x_{ij}=\left\{\begin{array}{rcl}\frac{1}{\sqrt {vol(C_j)}}  & & {v_i\in{C_j}\textrm{,} }\\0 & & {otherwise.}\end{array} \right.
\end{equation}
It is easy to verify $x_j^T(D-W)x_j=\frac {cut(C_j\textrm{,}\bar{C_j})}{vol(C_j)}$ and $X^TDX = E$, where $E$ is an identity matrix and the degree matrix
$D$ is defined as the diagonal matrix whose entry is $d_i=\sum_j^n w_{ij}$, degree of $v_i$. \par
Next, the unnormalized graph Laplacian $L$  is defined as follows:
\begin{equation}
\label{equ:unlap}
L=D-W.
\end{equation}
With matrices $X$ and $L$, the minimization problem in Eq. (\ref{equ:Ncut}) can be rewritten as the following problem:
\begin{equation}
\label{equ:minOne}
\begin{split}
    &\min\limits_{C}\quad Tr(X^TLX)\\
    &s.t. \quad X^TDX=E.
\end{split}
\end{equation}
Then, relaxing the discreteness condition and substituting $Y=D^\frac{1}{2}X$, the following relaxed problem is obtained :
\begin{equation}
\label{equ:minTwo}
\begin{split}
    &\min\limits_{Y\in{R^{n \times k}}} \quad Tr(Y^TL_NY) \\
    &s.t. \quad Y^TY=E\textrm{,}  \\
\end{split}
\end{equation}
where\\
\begin{equation}
\label{equ:norlap}
    L_N=D^{-\frac{1}{2}}(D-W)D^{-\frac{1}{2}}
\end{equation}
is a normalized graph Laplacian.
  Eq. (\ref{equ:minTwo}) is the standard form of a trace minimization problem. The Rayleigh-Ritz theorem \cite{handbook} tells us that its solution is the
  matrix whose columns are the first $k$ eigenvectors of matrix $L_N$ (By ``the first k eigenvectors" we refer to the eigenvectors corresponding to the k
  smallest eigenvalues).  Also, it is obvious that solution $X$ consists of the first $k$ generalized eigenvectors of $Lu = \lambda Du$ \cite{tutorial}.

 The algorithm of normalized spectral clustering by Shi and Malik \cite {shi}  is presented in Algorithm \ref{alg:KNSC}. Its computational complexity is
 $O(n^{\frac{3}{2}})$; its memory cost is $O(n)$.\par


\begin{algorithm}[h]
\caption{Normalized spectral clustering according to Shi and Malik \cite{shi}, \cite{Nascimento}}
\label{alg:KNSC}
\hspace*{0.02in} {\bf Input:}
The similarity matrix $W$ and the number of desired clusters $k$.

\begin{algorithmic}[1]
\State Find the first $k$ eigenvectors of the generalized eigensystem $Lu=\lambda Du$ and sort them in the columns of the matrix $U$. The $i$-th row of the
matrix $U$ will represent pixel  $v_i$ from image $I$.
\State Apply the $k$-means algorithm to matrix $U$ to find $k$ clusters $\pi =\{\pi_1\textrm{,} \pi_2\textrm{,} \ldots \textrm{,}  \pi_k\}$.
\State Form the final clusters assigning by clustering every node $v_i$, with $1\le i \le n$, into cluster $C_l$, if the $i$-th row of $U$ belongs to
$\pi_l$ in partition $\pi$.
\end{algorithmic}
\hspace*{0.02in} {\bf Output:}
The final clusters.
\end{algorithm}

\subsection{Quad-tree Decomposition}
\label{subsection:Pri-qp}

Quad-tree is a widely used tree data structure in the field of image segmentation \cite{Spann,Elsayed,Allen,Davatzikos,Weis}, and it is a spatial search
tree in which each internal node has exactly four child nodes. It is the two-dimensional analog of octrees.
 Quad-tree decomposition divides a square image into four equal-sized square blocks, and tests each block to see if it meets some criterion of
 homogeneity. If a block meets the criterion, it is not divided any further. Otherwise, it is subdivided again into four blocks. This process is repeated
 iteratively until each block meets the criterion. The final result includes multiple sizes of blocks.

   The typical criterion is as follows:
 \begin{equation}
\label{equ:quadtree}
\ var (\Omega) < t\textrm{,}
\end{equation}
where $\Omega$ represents an image block,  $var (\Omega)$ the variance of the pixel intensities of $\Omega$ and $t$ the threshold of quad-tree
decomposition.
Note that image $I$ can be divided into $\frac{1}{2}\log n$ levels at most \cite{Spann}. \figurename \ref{fig:quarterTree} shows the structure of a
quad-tree.\par

\begin{figure*}
\label{fig:quarterTree}
\centering
    \subfigure[]{
        \includegraphics[width=2in,height=2in]{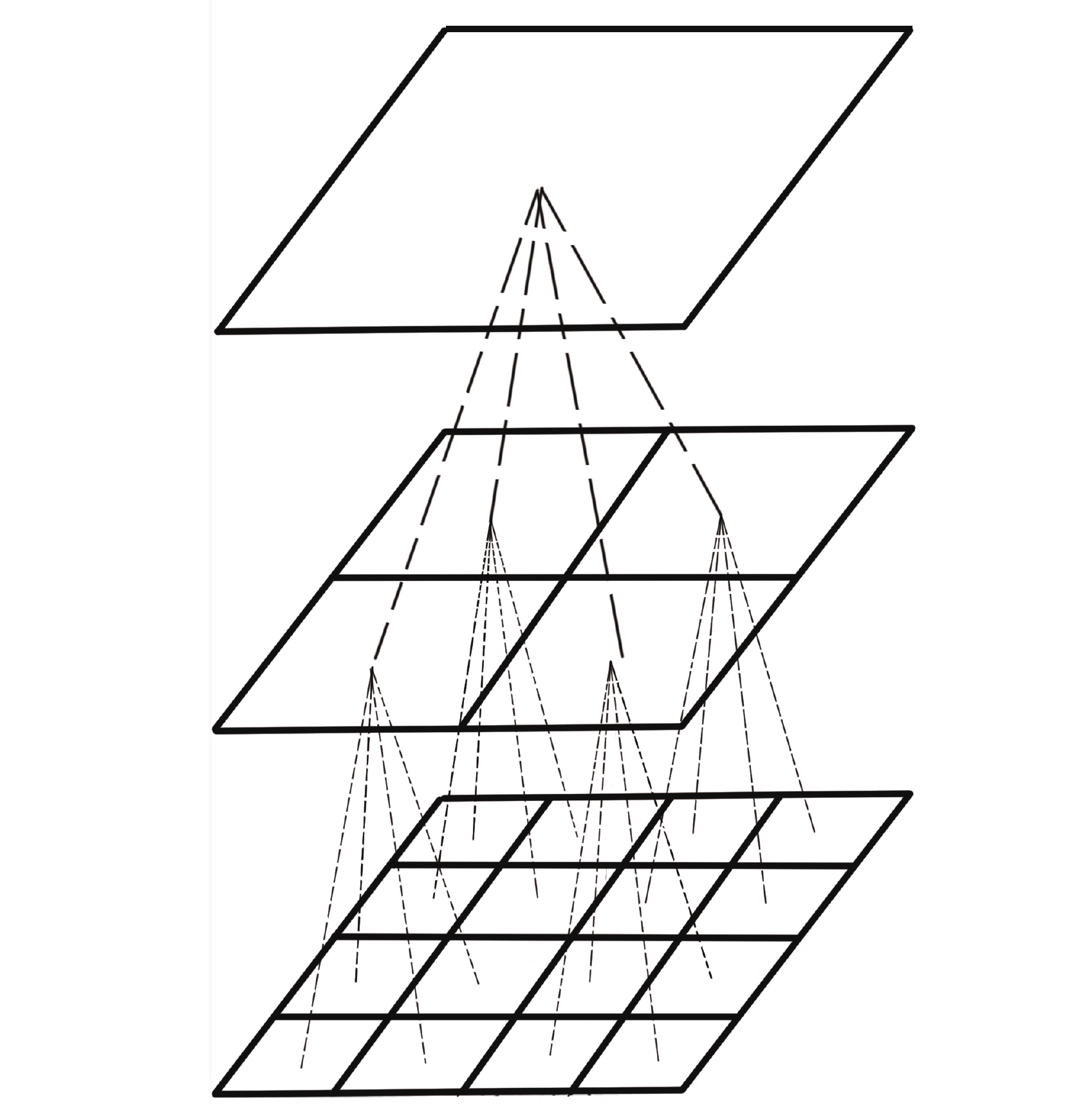}
    }
    \subfigure[]{
        \includegraphics[width=2in,height=2in]{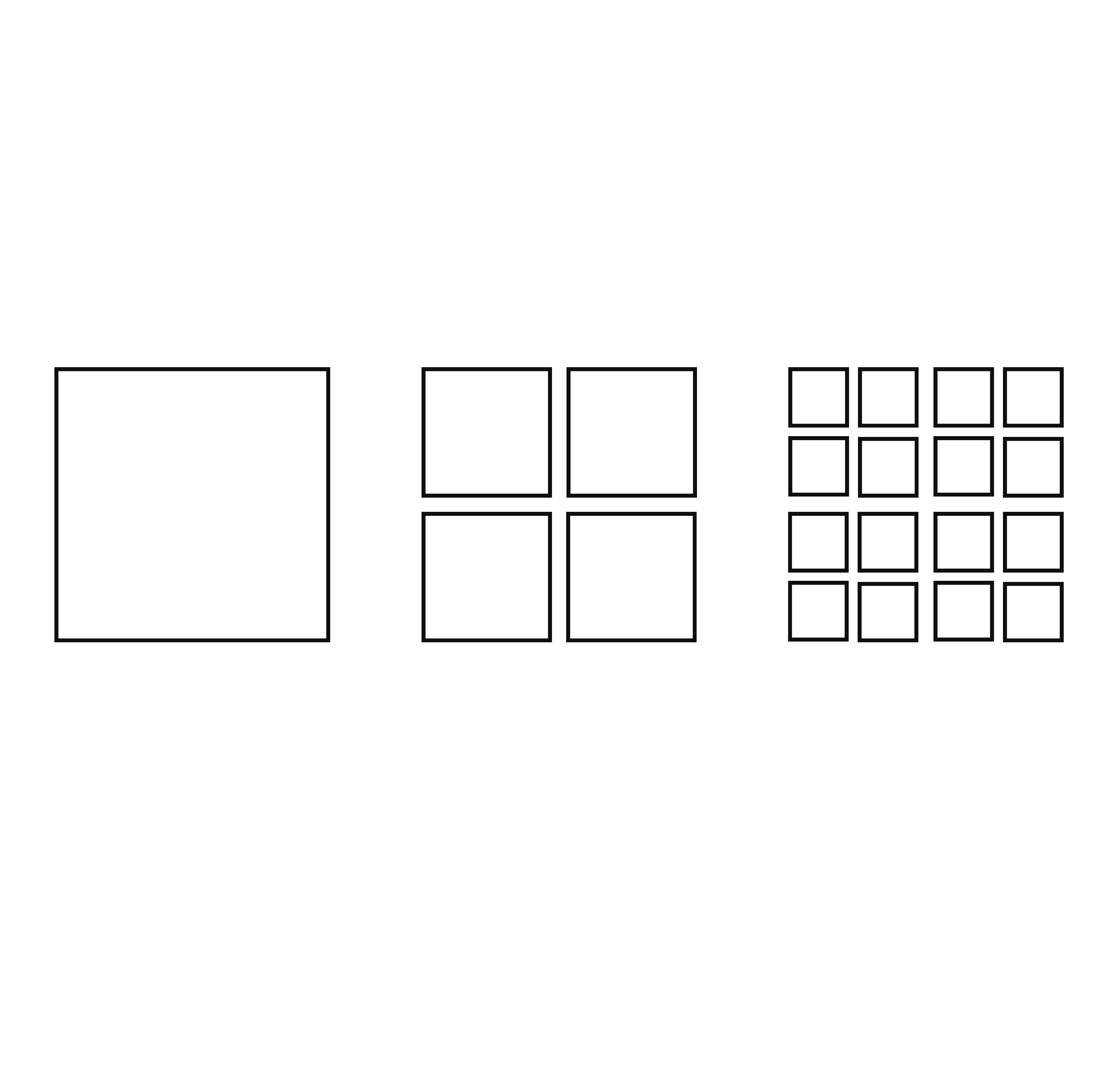}
    }\\
    \subfigure[]{
        \includegraphics[width=2in,height=2in]{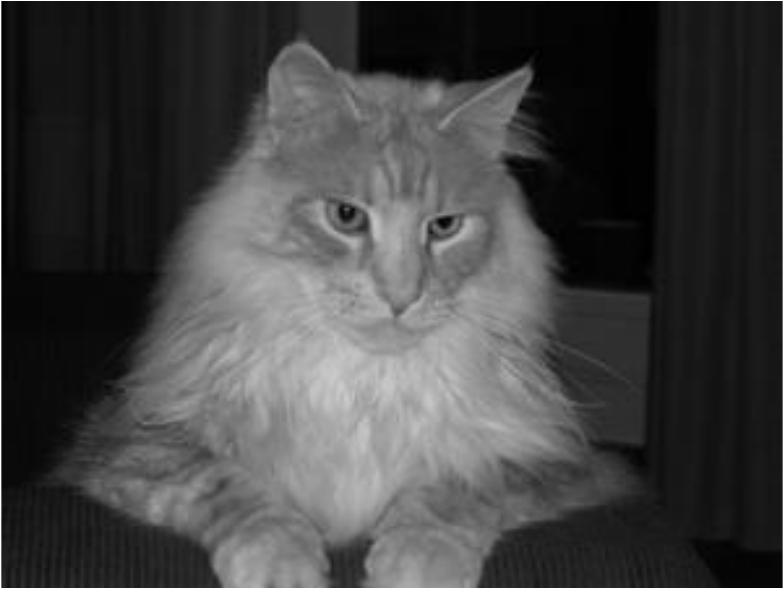}
    }
    \subfigure[]{
        \includegraphics[width=2in,height=2in]{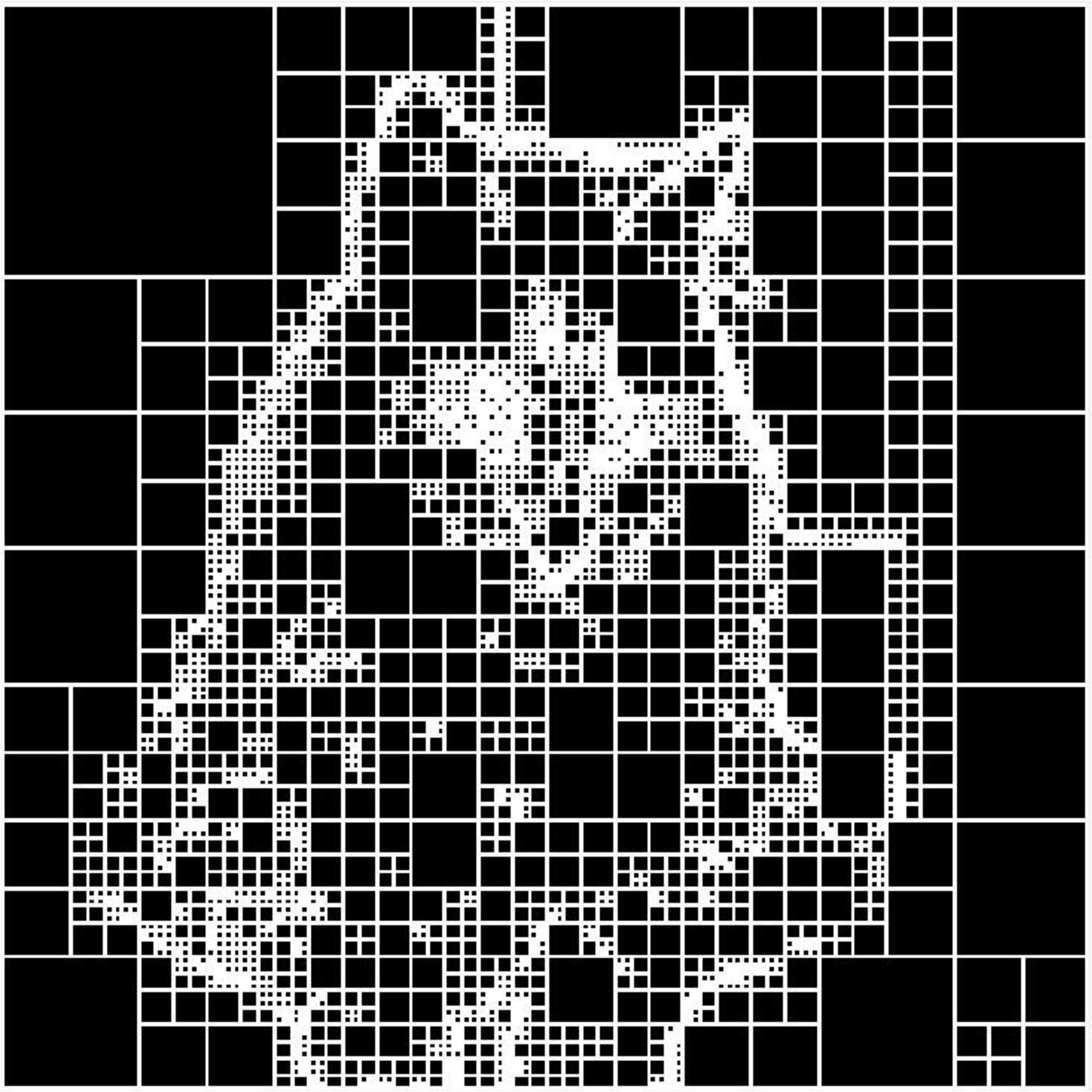}
    }
    \caption{ The structure of the quad-tree and the result of quad-tree decomposition. (a) the hierarchical structure of the quad-tree, (b)  the planar
    structure of the quad-tree, (c)  the image selected from Weizmann data set \cite{website}, (d) the quad-tree decomposition result of (c).}
    \label{fig:quarterTree}
\end{figure*}

\section{Methods}
\label{section:Method}
\subsection{Fast Spectral Clustering}
\label{subsection:Met-FSC}
In this section, we give a detailed introduction to FSC. It focuses on the spectral clustering at superpixel level. Suppose image $I$ is composed of $m$
superpixels, i.e., $I=\displaystyle\bigcup_{i=1}^{m}A_i$, where $A_i$ is the $i$-th superpixel and the number of superpixels $m$ is much smaller than the
number of pixels $n$. In FSC, we treat the leaf node blocks of the quad-tree as superpixels. Next, we start to solve the problem of spectral image segmentation
based on superpixels with FSC.  This problem is to divide the set of superpixels into $k$ clusters $B=\{B_1\textrm{,} ...\textrm{,} B_k\}$. Analogous to
Ncut, the problem is equivalent to the following minimization problem:
\begin{equation}
\label{equ:minFSC}
\min\limits_{B_1\textrm{,}B_2\textrm{,} ...\textrm{,} B_k} \quad FSC(B_1\textrm{,} B_2\textrm{,} ...\textrm{,} B_k)\textrm{,}
\end{equation}
where $FSC(B_1\textrm{,} B_2\textrm{,} ...\textrm{,} B_k)$ is defined as:
\begin{equation*}
\begin{aligned}
FSC(B_1\textrm{,} B_2\textrm{,} ...\textrm{,} B_k)=\displaystyle\sum_{i=1}^k\sum_{j=1\textrm{,}j\neq i}^k\frac{cut(B_i\textrm{,} B_j)}{vol(B_i)}\textrm{,}
\\
\end{aligned}
\end{equation*}
and
\begin{equation*}
\begin{split}
 &cut(B_i\textrm{,} B_j)=\displaystyle\sum_{A_i\in B_i}\sum_{A_j\in B_j}\frac{cut(A_i\textrm{,} A_j)}{\sqrt{|A_i||A_j|}}\textrm{,} \\
 &vol(B_i)=\displaystyle\sum_{A_j\in B_i}\sum_{z=1}^{m}\frac{cut(A_j\textrm{,} A_z)}{\sqrt{|A_j||A_z|}}\textrm{,} \\
\end{split}
\end{equation*}
where $|A_i|$  is the number of pixels in superpixel $A_i$, and $cut(A_i\textrm{,} A_j)=\displaystyle\sum_{v_i\in {A_i}\textrm{,} v_j\in A_j}w_{ij}$. In
order to solve the problem defined in Eq. (\ref{equ:minFSC}), the unnormalized graph Laplacian based on superpixels and the indicator vector of cluster
$B_j$ $(j = 1\textrm{,}2\textrm{,}...\textrm{,}k)$ are required. First, to construct the unnormalized graph Laplacian based on
superpixels, define the indicator vector of superpixel $A_j$ $(j=1\textrm{,} 2\textrm{,} ...\textrm{,} m)$ as $h_j=(h_{1j}\textrm{,}h_{2j}\textrm{,}
...\textrm{,} h_{nj})^T$ by
\begin{equation}
\label{equ:indicator}
\begin{split}
h_{ij}=\left\{ \begin{array}{rcl} \frac{1}{\sqrt{|A_j|}} & & v_i \in A_j\textrm{,}\\ 0 \qquad{} & & otherwise \end{array}\right. \\
\end{split}
\end{equation}
$\qquad{}\qquad{}\qquad{}\qquad{}\qquad{}(i=1\textrm{,} 2\textrm{,} ...\textrm{,} n)\textrm{,}  $
\vspace{8pt}
\\ where $|A_j|$ is the number of pixels in superpixel $A_j$. Then we construct matrix $H\in \mathbb{R}^{n\times m}$ whose columns are the indicator
vectors of $A_j$ $(j = 1\textrm{,}...\textrm{,}m)$. Matrix $H$ is a transformation mapping the pixel space to the superpixel space. It is easy to observe
that matrix $H$ is a columns orthogonal matrix.

With matrices $H$ and $W$, we obtain the following similarity matrix $\widetilde{W}$ based on superpixels:
\begin{equation}
\label{equ:Wf}
\begin{aligned}
\widetilde{W}&=H^{T}WH\\&=\left[ \begin{array}{ccc} \frac{cut(A_1\textrm{,} A_1)}{|A_1|} & \cdots & \frac{cut(A_1\textrm{,}
A_m)}{\sqrt{|A_1||A_m|}}\\\vdots & \ddots & \vdots\\  \frac{cut(A_m\textrm{,} A_1)}{\sqrt{|A_m||A_1|}} & \cdots & \frac{cut(A_m\textrm{,}A_m)}{|A_m|}
\end{array} \right ].\\
\end{aligned}
\end{equation}

Now we are able to define the degree matrix $\widetilde D$ based on superpixels:
\begin{equation}
\label{equ:Df}
\begin{aligned}
\widetilde D&=\left[ \begin{array}{ccc}  d_{1} &  & \\  & \ddots &  \\   &   & d_{m} \end{array} \right ]\textrm{,} \\
d_{i}&=\displaystyle\sum_{j=1}^m\widetilde{W}_{ij}.
\end{aligned}
\end{equation}

With matrices $\widetilde{W}$ and $\widetilde D$, the unnormalized graph Laplacian based on superpixels $\widetilde L$ is defined as follows:
\begin{equation}
\label{equ:Lf}
\begin{aligned}
\widetilde L=\widetilde D-\widetilde W.
\end{aligned}
\end{equation}

Next, we define the indicator vector $g_j=(g_{1j}\textrm{,} g_{2j}\textrm{,} ...\textrm{,} g_{mj})^T$ of $B_j\;(j =
1\textrm{,}2\textrm{,}...\textrm{,}k)$:

\begin{equation}
\label{equ:g}
\begin{split}
g_{ij}=\left\{ \begin{array}{rcl} \frac{1}{\sqrt{vol(B_j)}} & & A_i \in B_j\textrm{,}\\ 0 \qquad{} & & otherwise \end{array} \right.\\
\end{split}
\end{equation}
$\qquad{}\qquad{}\qquad{}\qquad{}\qquad{}(i=1\textrm{,} 2\textrm{,} ...\textrm{,} n)\textrm{,}  $
\vspace{8pt}
\\It is easy to obtain the following equations:

\begin{equation*}
\begin{aligned}
g_j^T\widetilde D g_j&=\frac{\displaystyle\sum_{A_i\in B_j}\sum_{z=1}^{m}\frac{cut(A_i\textrm{,} A_z)}{\sqrt{|A_i||A_z|}}}{vol(B_j)}\\
&=1\textrm{,}
\end{aligned}
\end{equation*}

\begin{equation*}
\begin{aligned}
g_j^T\widetilde Lg_j &=\displaystyle\sum_{i=1\textrm{,} i\ne j}^k\sum_{A_a\in B_i}\sum_{A_b\in B_j}\frac{cut(A_a\textrm{,}
A_b)}{vol(B_j)\sqrt{|A_a||A_b|}}\\
 &=\displaystyle\sum_{i=1\textrm{,} i\neq j}^k\frac{cut(B_j\textrm{,} B_i)}{vol(B_j)}.
\end{aligned}
\end{equation*}

Then, construct matrix $G\in \mathbb{R}^{m\times k}$ whose columns are the indicator vectors of $B_j\;(j = 1\textrm{,}2\textrm{,}...\textrm{,}k)$.

Therefore, the minimization problem of Eq. (\ref{equ:minFSC}) can be rewritten as:
\begin{equation}
\label{equ:newMin}
\begin{split}
&\min\limits_B \quad Tr(G^T\widetilde LG) \\
&s.t. \quad G^T\widetilde DG=E\textrm{,}
\end{split}
\end{equation}
where $Tr$ denotes the trace of a matrix.
Relax the above problem by allowing the entries of matrix $G$ to take arbitrary real values. Substitute $P=\widetilde D^{\frac{1}{2}}G$. Now we obtain the
following relaxed problem:
\begin{equation}
\label{equ:minNew2}
\begin{split}
&\min\limits_{P\in R^{m\times k}}Tr(P^T\widetilde L_NP) \\
&s.t. \quad P^TP=E\textrm{,}
\end{split}
\end{equation}

where
\begin{equation*}
\begin{aligned}
\widetilde L_N=\widetilde D^{-\frac{1}{2}}\widetilde L\widetilde D^{-\frac{1}{2}}.
\end{aligned}
\end{equation*}
%

 The minimization problem of Eq. (\ref{equ:minNew2}) is the standard form of a trace minimization problem. The Rayleigh-Ritz theorem \cite{handbook} tells
 us that its solution is the first k eigenvectors of Laplacian $\widetilde L_N$. Obviously, $\widetilde L_N$ is $m\times m$ in size, and is sparse.
 Therefore, the computational complexity of solving  the eigenvectors of $\widetilde L_N$ is much lower than that of solving the eigenvector of  Laplacian
 matrix whose size is $n \times n$ in Ncut. In section \ref{subsection:Met-running time}, we will analyze the complexity of FSC in detail. To obtain the
 matrix with clustering information based on pixels, we convert $G$ based on superpixels to $G_{p}$ based on pixels:
\begin{equation}
G_{p}=HG\textrm{,}
\label{equ:trans}
\end{equation}
where $H$ is defined in Eq. (\ref{equ:indicator}).
Then we treat each row of $G_{p}$ as a point $\in \mathbb{R}^k$ and cluster all of the points into $k$ clusters via the Fuzzy C-means algorithm to obtain
the clustering result based on pixels.

\par

The details of FSC are given in Algorithm \ref{alg:fsc}.

\begin{algorithm}[h]
\caption{Fast Spectral Clustering (FSC)}
\label{alg:fsc}
\hspace*{0.02in} {\bf Input:}
Image $I$, the number of clusters $k$, the number of superpixels $m$.
\begin{algorithmic}[1]
\State Obtain superpixels $A=\{A_1\textrm{,} A_2\textrm{,} ...\textrm{,} A_m\}$ by quad-tree decomposition.
\State Form the superpixel-based similarity matrix $\widetilde W\in \mathbb{R}^{m\times m}$ and superpixel-based degree matrix $\widetilde D$ as in Eq. (\ref{equ:Wf}), (\ref{equ:Df}).
\State Compute the superpixel-based normalized graph Laplacian $\widetilde L_N=\widetilde D^{-\frac{1}{2}}(\widetilde D - \widetilde W)\widetilde D^{-\frac{1}{2}}$.
\State Compute the eigenvectors $t^1\textrm{,} t^2\textrm{,} ...\textrm{,} t^k$ that the first k eigenvalues of matrix $\widetilde L_N$ correspond to.
\State Form the indicator matrix $T=[t^1\textrm{,} t^2\textrm{,} ...\textrm{,} t^k]\in \mathbb{R}^{m \times k}$.
\State Compute matrix $G=\widetilde D^{-\frac{1}{2}}T$.

\State Convert matrix $G$ to $G_{p}$  by Eq. (\ref{equ:trans}).
\State Treat each row of $G_{p}$ as a point $\in \mathbb{R}^k$ and cluster all of the points into $k$ clusters via the Fuzzy C-means algorithm to obtain the clustering result based on pixels.
\end{algorithmic}
\hspace*{0.02in} {\bf Output:}
$k$ clusters.
\end{algorithm}

\subsection{Multiscale Fast Spectral Clustering(MFSC)}
\label{subsection:Met-MFSC}

Though the number of superpixels is smaller than that of pixels $(m<n)$, the computational complexity of FSC is still high on complex images. To address
the problem, we propose the Multiscale Fast Spectral Clustering (MFSC).\par
First, we construct the hierarchical structure of image $I$ by quad-tree decomposition. Second,  we construct a superpixel-based similarity matrix
according to Eq. (\ref{equ:Wf}). Third, we treat the result of segmenting the child nodes at the fine level as the superpixels of the parent nodes at the
coarse level, and obtain the segmentation result at the coarse level by using FSC on these superpixels. The process is repeated  level by level until
the coarsest level is reached and the segmentation results of the whole image is finally obtained.
Next, we will expound on MFSC.

Suppose that $S^{l-1}$ a node at level $l-1$, contains superpixels $A_{S^{l-1}}=\{A_{i_1}\textrm{,} A_{i_2}\textrm{,} ...\textrm{,} A_{i_s}\}$,
 and its four child nodes are $S_{1}^l$, $S_{2}^l$, $S_{3}^l$ and $S_{4}^l$ which consist of the following superpixels respectively:

\begin{equation}
\begin{aligned}
S_{1}^l&=\{A^{1}_1\textrm{,} A^{1}_2\textrm{,} ...\textrm{,} A^{1}_{k_1}\}\textrm{,}\\
S_{2}^l&=\{A^{2}_1\textrm{,} A^{2}_2\textrm{,} ...\textrm{,} A^{2}_{k_2}\}\textrm{,}\\
S_{3}^l&=\{A^{3}_1\textrm{,} A^{3}_2\textrm{,} ...\textrm{,} A^{3}_{k_3}\}\textrm{,}\\
S_{4}^l&=\{A^{4}_1\textrm{,} A^{4}_2\textrm{,} ...\textrm{,} A^{4}_{k_4}\}\textrm{,}
\end{aligned}
\end{equation}

where $S^{l-1}=\displaystyle\bigcup_{i\in \{1\textrm{,} 2\textrm{,} 3\textrm{,} 4\}}S_{i}^l$. \par
Fig. \ref{fig:granode} shows the relationship between parent node $S^{l-1}$ and its four child nodes $S_{1}^l$, $S_{2}^l$, $S_{3}^l$, $S_{4}^l$.
\begin{figure}
\label{fig:granode}
\centering
        \includegraphics[width=1.9in,height=1.5in]{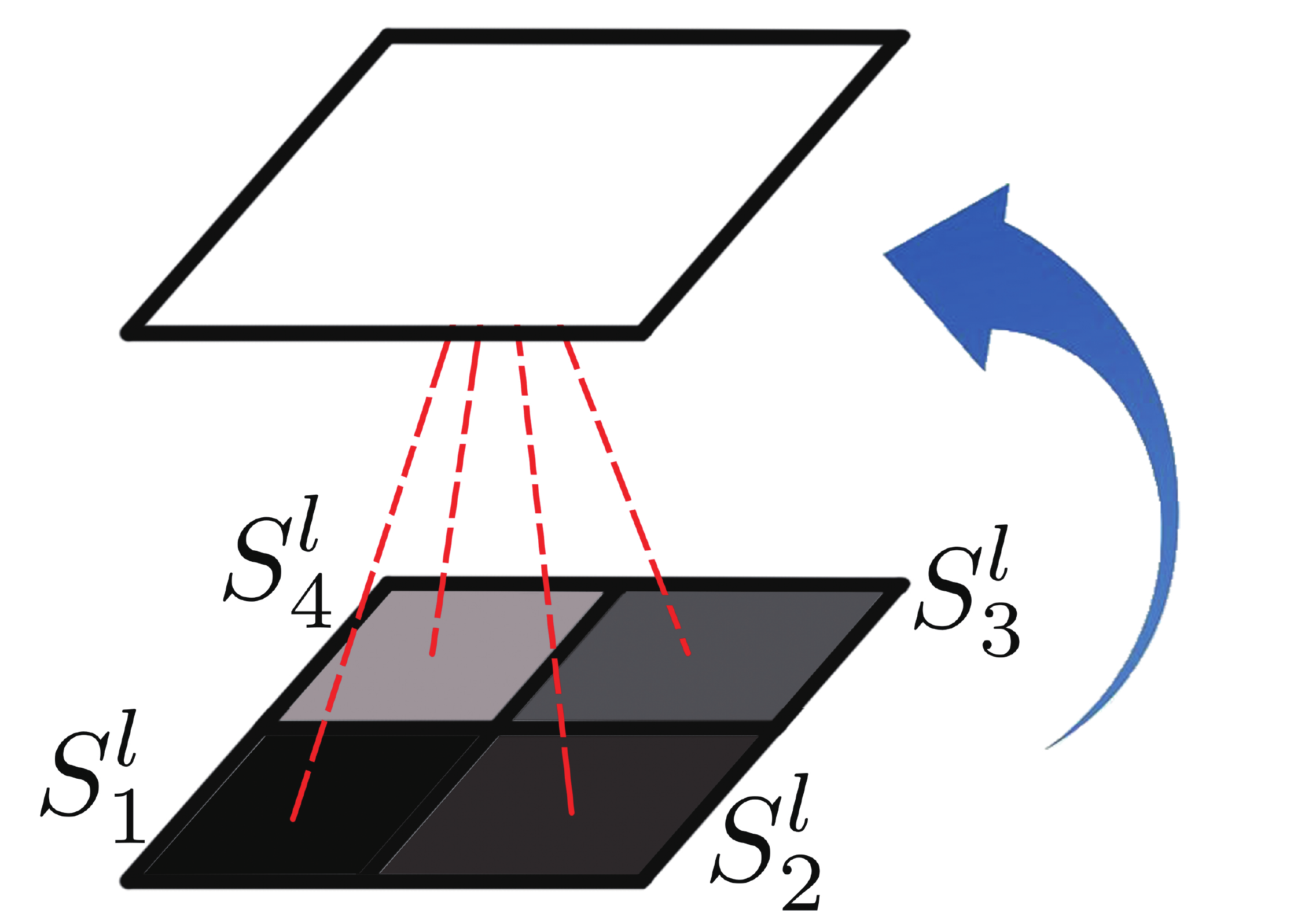}
\caption{The relationship between parent node $S^{l-1}$ and its child nodes $S_{ul}^l$, $S_{ur}^l$, $S_{dl}^l$ and $S_{dr}^l$.}
    \label{fig:granode}
\end{figure}

Suppose that we have obtained the clustering result $C^{l}_{1}=\{(C^{l}_{1})_1\textrm{,} ...\textrm{,} (C^{l}_{1})_o\}$ of $S^{l}_{1}$ via FSC, where $o$
($o\leq k_1$) represents the number of the clusters in $C^{l}_{1}$. We use $Q_{1}^l$, $Q_{2}^l$, $Q_{3}^l$, $Q_{4}^l$ to represent the indicator
matrices of the clustering results of the  four child nodes.
The entries of matrix $Q_{1}^l$ is defined as:

\begin{equation}
\label{equ:indicatorc}
\begin{split}
Q^l_{1}(i\textrm{,} j)=\left\{ \begin{array}{rcl} \frac{1}{\sqrt{|(C^{l}_{1})_j|}} & & A^{1}_{i} \in (C^{l}_{1})_j\textrm{,} \\ 0 & & otherwise
\end{array} \right.\\
(i=1\textrm{,} 2\textrm{,} ...\textrm{,} k_{1}\;and\;j=1\textrm{,} 2\textrm{,} ...\textrm{,} o)\textrm{,}
\end{split}
 \end{equation}
where $|(C^{l}_{1})_j|$ is the number of superpixels in cluster $(C^{l}_{1})_j$. Similarly, we can define the indicator matrices of the other child
nodes in the same way.
In particular, suppose that MFSC starts from  level $l_{init}$ of the quad-tree. Then each block below level $l_{init}$ is treated as a cluster.  Hence its
indicator matrix is an identity matrix.
 With the indicator matrices of the four child nodes, we define matrix
 \begin{equation}
\label{equ:Ql}
\begin{aligned}
&Q^{l}=\left[ \begin{array}{cccc}  Q_{1}^l &  & & \\  & Q_{2}^l &  & \\   &   & Q_{3}^l & \\  &   &  & Q_{4}^l \end{array} \right ]_{s\times
e}\textrm{,} \\
\end{aligned}
\end{equation}
where $s$ is the number of rows in  $Q^{l}$ and is equal to the number of superpixels in node $S^{l-1}$;
  $e$ is the number of the columns of $Q^{l}$ and is equal to the number of the clusters of the four child nodes in total.\par
 Next,  cluster  set $C^{l}=\displaystyle\bigcup_{i\in \{1\textrm{,} 2\textrm{,} 3\textrm{,} 4\}}C^{l}_i$ by  FSC. First, the
 similarity matrix based on superpixel set $C^l$ is defined as:
 \begin{equation}
 \label{equ:w_F_S}
\begin{aligned}
 \widetilde{W}_{S^{l-1}}=(Q^{l})^{T}W_{S^{l-1}}Q^{l}\textrm{,}
\end{aligned}
\end{equation}
 where $W_{S^{l-1}}$ is the similarity matrix based on superpixels of node $S^{l-1}$ and the $s$-order submatrix of $\widetilde{W}$ defined in Eq.
 (\ref{equ:Wf}).

Also, we can define the degree matrix $\widetilde{D}_{S^{l-1}}$ based on the set of superpixels $C^{l}$ of node $S^{l-1}$ as follows:
\begin{equation}
\label{equ:Dsf}
\begin{aligned}
&\widetilde{D}_{S^{l-1}}=\left[ \begin{array}{ccc}  \widetilde{d}_{1} &  & \\  & \ddots &  \\   &   & \widetilde{d}_e \end{array} \right ]\textrm{,} \\
&  \widetilde{d}_{i}=\displaystyle\sum_{j=1}^e\widetilde{W}_{S^{l-1}}(i\textrm{,}j).
\end{aligned}
\end{equation}
With matrices $\widetilde{W}_{S^{l-1}}$ and $\widetilde{D}_{S^{l-1}}$, we define the normalized graph Laplacian matrix $\widetilde{L}_{S^{l-1}}$ based on
the set of superpixels $C^{l}$ of node $S^{l-1}$ as follows:
\begin{equation}
\label{equ:Lsf}
\begin{aligned}
\widetilde{L}_{S^{l-1}}=(\widetilde{D}_{S^{l-1}})^{-\frac{1}{2}}(\widetilde{D}_{S^{l-1}}-\widetilde{W}_{S^{l-1}})(\widetilde{D}_{S^{l-1}})^{-\frac{1}{2}}.
\end{aligned}
\end{equation}

Our aim is to further cluster the the set of superpixels $C^l$ into $k$ clusters $B=\{B_1\textrm{,}  B_2\textrm{,}  \ldots \textrm{,}
B_k\}$.
 Analogous to FSC, the clustering result can be obtained by solving the following minimization problem:
 \begin{equation}
\label{equ:weTwo}
\begin{aligned}
 \displaystyle& \min_{\widetilde{Q}^{l-1}\in R^{e \times k}} Tr((\widetilde{Q}^{l-1})^T\widetilde{L}_{S^{l-1}}\widetilde{Q}^{l-1})\\
 &s.t. \quad (\widetilde{Q}^{l-1})^T\widetilde{Q}^{l-1}=E\textrm{,}
\end{aligned}
\end{equation}
where $\widetilde{Q}^{l-1}$ is the indicator matrix of clusters $B$ based on the set of superpixels $C^l$.

In order to obtain the indicator matrix based on superpixels $A_{S^{l-1}}$, we convert the indicator matrix $\widetilde{Q}^{l-1}$ to matrix $Q^{l-1}$:
\begin{equation}
\label{equ:eTwo}
\begin{aligned}
Q^{l-1}=Q^{l}\widetilde{Q}^{l-1}.
\end{aligned}
\end{equation}
Here, we omit the Fuzzy C-means and treat matrix $Q^{l-1}$ as the indicator matrix of the clustering result of the node $S^{l-1}$.  \par
According to Section \ref{subsection:Met-FSC}, the solution of the problem of Eq. (\ref{equ:weTwo}) is the first $k$ eigenvectors of matrix
$\widetilde{L}_{S^{l-1}}$. It is obvious that the size of matrix $\widetilde{L}_{S^{l-1}}$ is much smaller than that of matrix $\widetilde L_N$.
Therefore, the computational complexity is further reduced.\par
Repeat the above procedures along the quad-tree structure from the bottom fine level to the top coarse level until the final superpixel-based indicator
matrix $Q^1$ is obtained.
We convert matrix $C_{sup}=\widetilde D^{-\frac{1}{2}}Q^1$ to $C_{p}$ with the clustering information based on pixels:
\begin{equation}
C_{p}=HC_{sup}\textrm{,}
\label{equ:trans1}
\end{equation}
where $H$ is defined in Eq. (\ref{equ:indicator}). Then we treat each row of $C_{p}$  as a point $\in \mathbb{R}^k$ and cluster all of the points into $k$
clusters via the Fuzzy C-means algorithm to obtain the clustering result based on pixels.\par
The details of MFSC are given in Algorithm \ref{alg:msfsc}.

\begin{algorithm}[h]
\caption{Multiscale Fast Spectral Clustering (MFSC)}
\label{alg:msfsc}
\hspace*{0.02in} {\bf Input:}
image $I$, the number of clusters $k$, the number of superpixels $m$, the start level $l_{init}$.

\begin{algorithmic}[1]
\State Decompose image $I$ by quad-tree decomposition to obtain superpixels $A$ and quad-tree $T$.
\State Form the superpixel-based similarity matrix $\widetilde W\in \mathbb{R}^{m\times m}$ by Eq. (\ref{equ:Wf}).
\State Set $l=l_{init}$;
\While {$l\geq 1$}
\For {each node at level $l$}
\If {$l=l_{init}$}
\State set the indicator matrix of the clustering result of current node $Q^{l}=E$;

\Else
\State compute $Q^{l}$ according to Eq. (\ref{equ:eTwo});
\EndIf
\EndFor
\State $l \leftarrow l-1$;
\EndWhile

\State Compute matrix $C_{p}$ with the clustering information based on pixels by Eq. (\ref{equ:trans1}).

\State Treat each row of $C_{p}$ as a point in $\mathbb{R}^k$ and cluster all of the points into $k$ clusters via the Fuzzy C-means algorithm to obtain
the final clusters of image $I$.

\end{algorithmic}

\hspace*{0.02in} {\bf Output:}
$k$ clusters
\end{algorithm}

\subsection{Computation and Memory Cost Analysis}
\label{subsection:Met-running time}

The computational complexity of the proposed FSC consists of the time of quad-tree decomposition, construction of the superpixel-based similarity matrix
and  eigen-decomposition of the superpixel-based Laplacian matrix. Their computational complexities are $O(n\log n)$, $O(m)$ and $O(m^\frac{3}{2})$,
respectively, where $n$, $m$ are the number of pixels and that of superpixels respectively. Hence the computational complexity of FSC is $O(n\log
n)+O(m)+O(m^\frac{3}{2})$. By contrast, Ncut takes $O(n)+O(n^{\frac{3}{2}})$, where $n>m$. The computational complexity of the proposed MFSC consists of
the following parts: it takes $O(1)$ to solve the first $k$ eigenvectors of matrix $\widetilde{L}_{S^{l-1}}$, $O(n)$ to compute matrix
$\widetilde{L}_{S^{l-1}}$ and $O(n)$ to perform quad-tree decomposition at each level. Hence the method takes $O(n\log{n})$ in total.\par
The two methods that we propose need to store the superpixel-based similarity matrix  $\widetilde{W}$, which counts for the storage of $O(m)$ real-valued
numbers since $\widetilde{W}$ is sparse. The memory cost of the two algorithms is much less than that of Ncut $O(n)$.

\ifCLASSOPTIONcaptionsoff
  \newpage
\fi

\section{Experiment and analysis}
\label{section:Experiment}

In this section, we test MFSC by doing a series of experiments on Weizmann data set, an image data set. To evaluate the accuracy of our method, the performances and results of Ncut on the same data set are recorded for comparison. To test the performance of the algorithm on images of different sizes, we scale the images into 3 different sizes ($128\times128$, $256\times256$, $512\times512$). Although the images will be distorted slightly after scaling, the comparison of segmentation results will not be affected. The following subsections describe the details of the experiments and results.\par

\subsection{Single-object Sample Images and Parameter Settings}

\begin{figure}[H]
\centering
    \subfigure[]{
        \includegraphics[width=1.5in,height=1.5in]{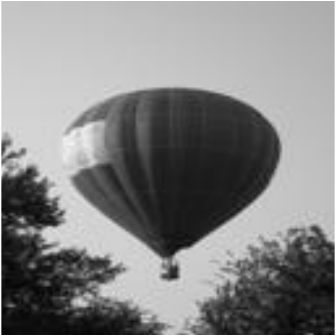}
    }
    \subfigure[]{
        \includegraphics[width=1.5in,height=1.5in]{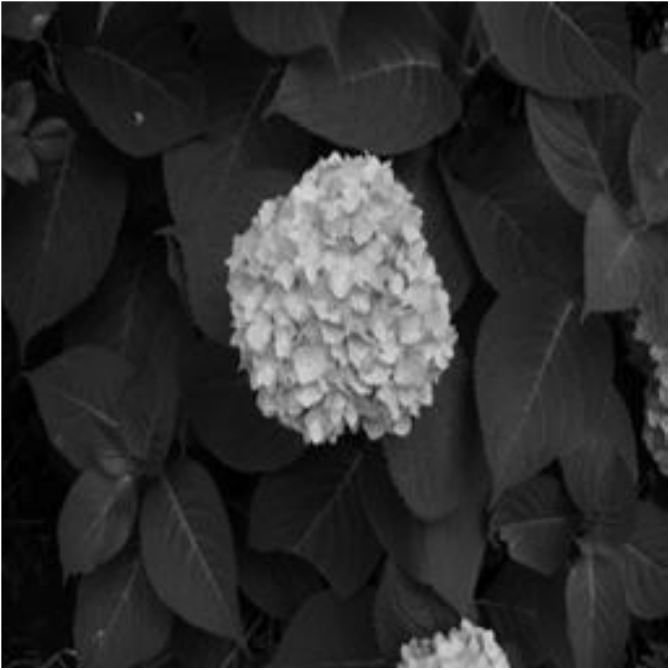}
    }
    \subfigure[]{
        \includegraphics[width=1.5in,height=1.5in]{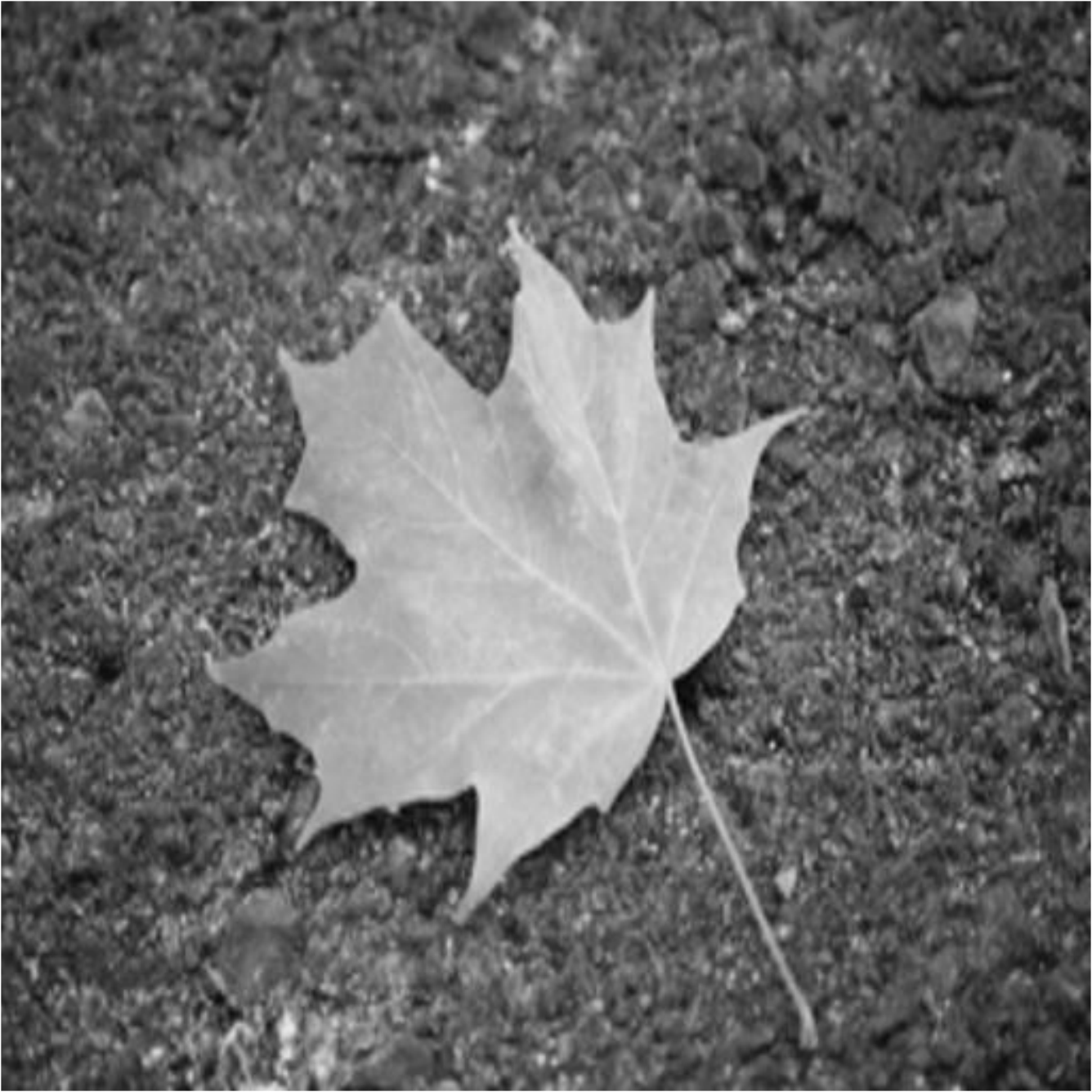}
    }
    \subfigure[]{
        \includegraphics[width=1.5in,height=1.5in]{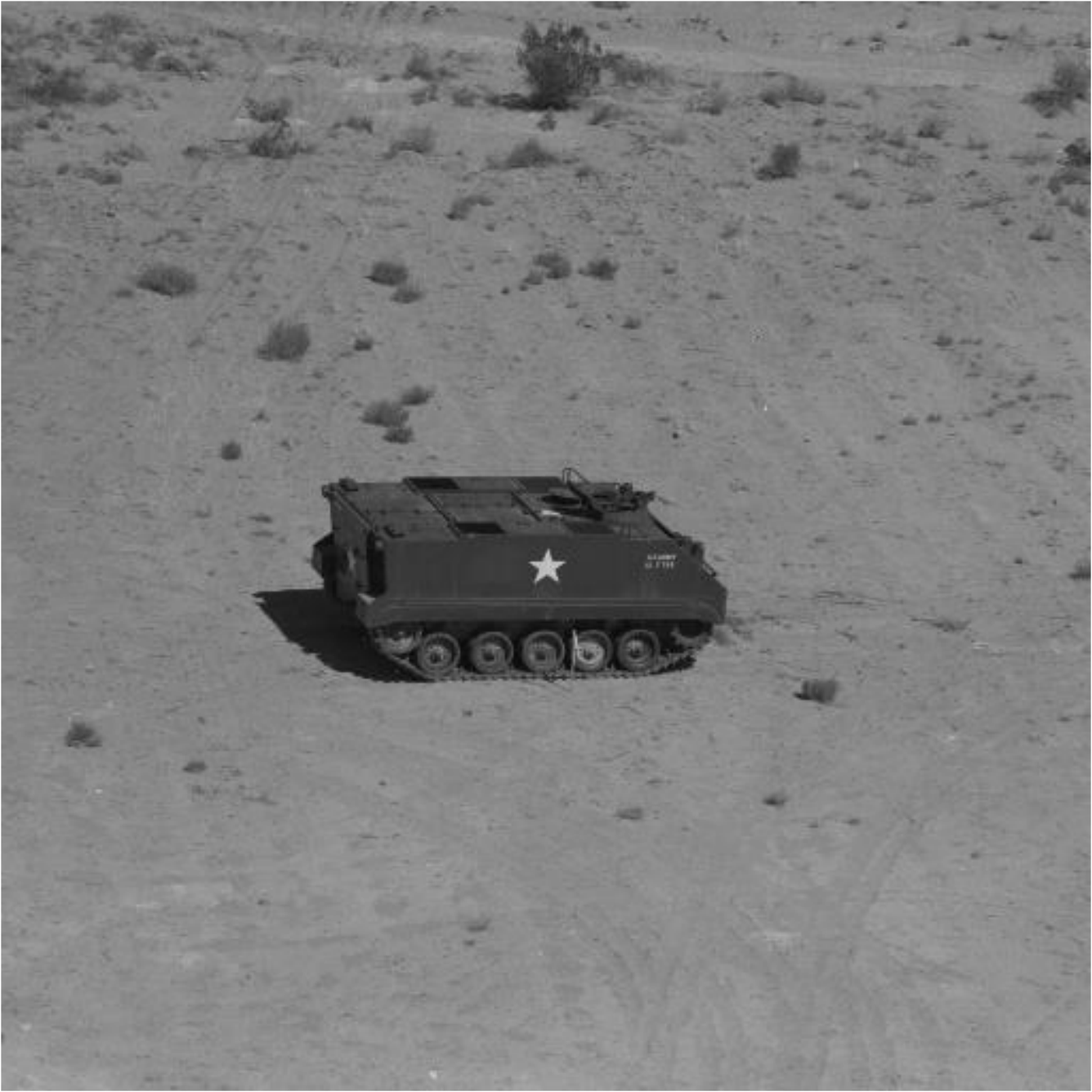}
    }
    \caption{Images used in the experiment. (a) HotAirBalloon ($128 \times 128$). (b) Nitpix ($256 \times 256$). (c) Leafpav72 ($512 \times 512$). (d) Tank ($512 \times 512$).}
    \label{fig:four Image}
\end{figure}

To show the segmentation results of MFSC and Ncut on single-object images, we select four sample images displayed in \figurename \ref{fig:four Image}. $HotAirBalloon$, $Nitpix$ and $Leafpav72$ are selected from Weizmann data set \cite{website}, whereas $Tank$ is selected from the database of University of Southern California \cite{website2}.\par

1) $HotAirBalloon$. The original size of the image is $300\times420$. We scale it to $128\times128$ pixels. We select the following parameters for MFSC: superpixel connection radius $R$ = 40 (Two superpixels are connected in a graph if their center pixels are within distance $R$), threshold of quad-tree decomposition $t = 10$, parameters of $\sigma_I = 8$, $\sigma_x = 4$, $\sigma_c = 0.2$ and $\alpha = 0.45$ for constructing the similarity matrix. The corresponding parameters of Ncut are set as: pixel connection radius of Ncut $r = 20$, $\sigma_c = 0.1$. \par

2) $Nitpix$. The original size of the image is $300\times225$. We scale it to $256\times256$ pixels. We select the following parameters for MFSC: $R = 40$, $t = 12$, $\sigma_I = 8$, $\sigma_x = 4$, $\sigma_c = 0.1$, $\alpha = 0.45$. The corresponding parameters of Ncut are: $r = 15$, $\sigma_c = 0.1$.\par

3) $Leafpav72$. The original size of the image is $300\times203$. We scale it to $512\times512$ pixels. We select the following parameters for MFSC: $R = 80$, $t = 12$, $\sigma_I = 8$, $\sigma_x = 4$, $\sigma_c = 0.09$, $\alpha = 0.45$. The corresponding parameters of Ncut are: $r = 10$, $\sigma_c = 0.1$.\par

4) $Tank$. The original size of the image is $512 \times 512$. We select the following parameters for MFSC: $R = 40$, $t = 7$, $\sigma_I = 0.12$, $\sigma_x = 0.12$, $\sigma_c = 0.1$, $\alpha = 0.1$. The corresponding parameters of Ncut are: $r = 10$, $\sigma_c = 0.1$.\par

\subsection{Two-object Sample Images and Parameter Settings}
\begin{figure}[H]
\centering
    \subfigure[]{
        \includegraphics[width=1.5in,height=1.5in]{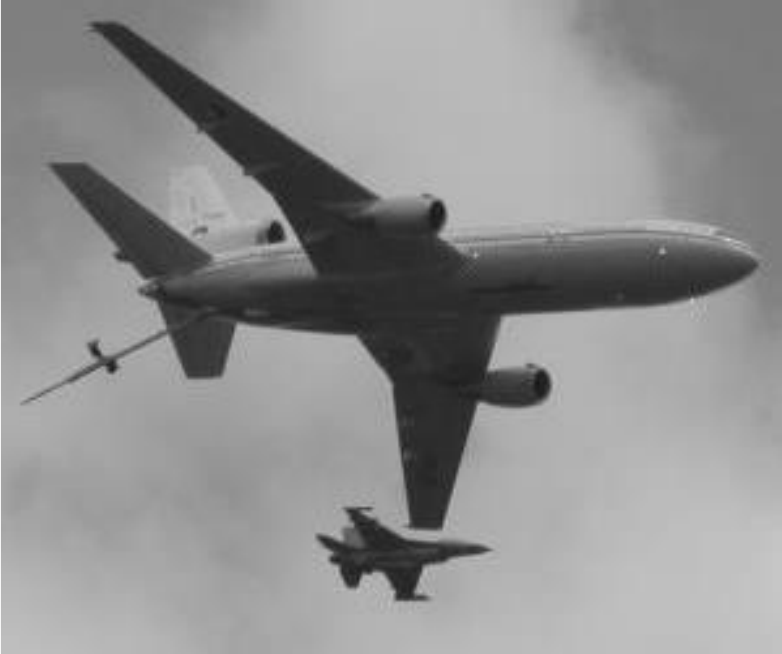}
    }
    \subfigure[]{
        \includegraphics[width=1.5in,height=1.5in]{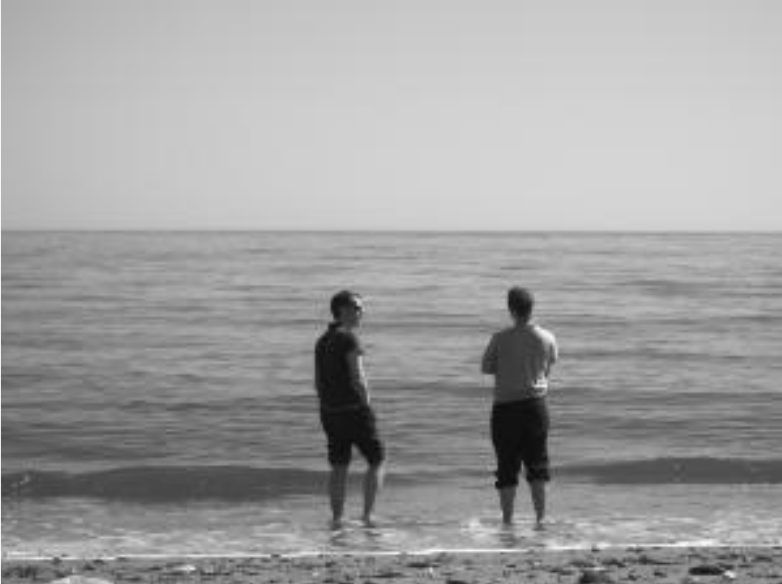}
    }
    \subfigure[]{
        \includegraphics[width=1.5in,height=1.5in]{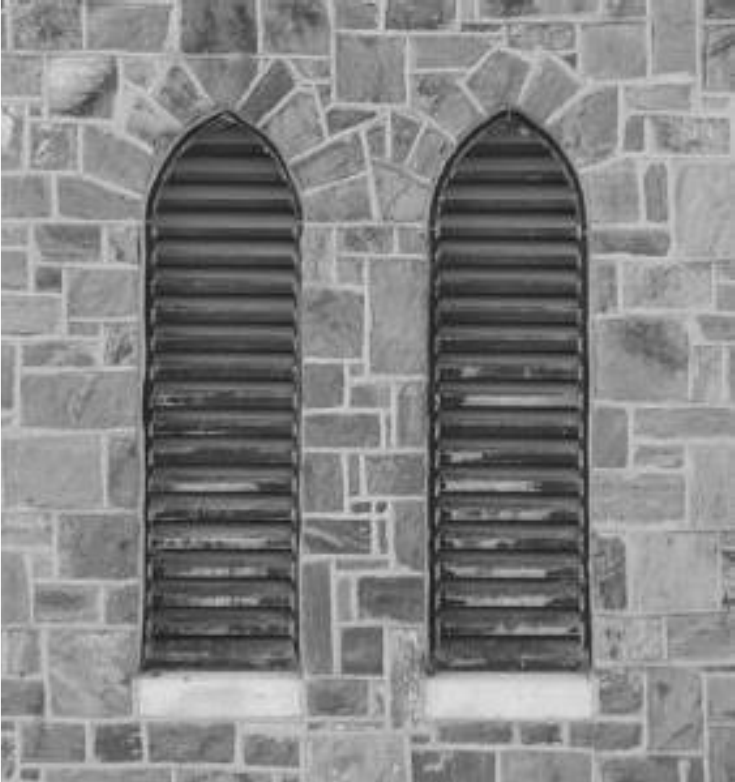}
    }
    \subfigure[]{
        \includegraphics[width=1.5in,height=1.5in]{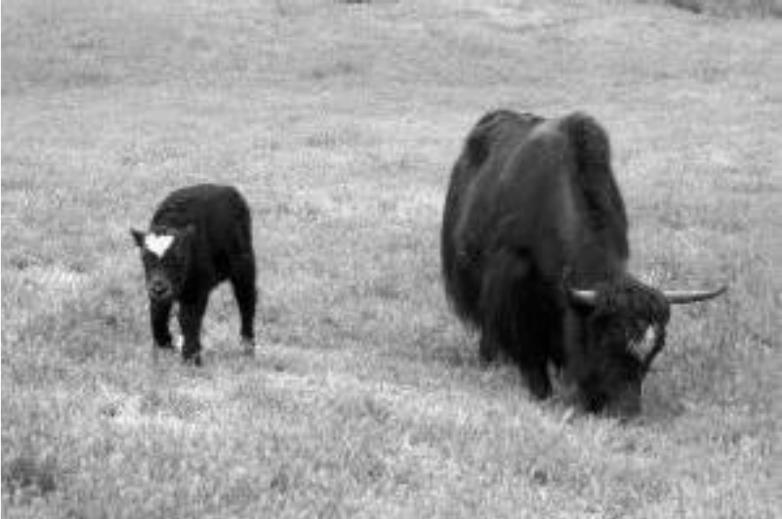}
    }
    \caption{Images used in the experiment. (a) Plane ($256 \times 256$). (b) Imgp1883 ($256 \times 256$). (c) DualWindows ($256 \times 256$). (d) Yack1 ($256 \times 256$).}
    \label{fig:object image}
\end{figure}

For multi-object images, we select four sample images displayed in \figurename \ref{fig:object image}. $Plane$, $Imgp1883$, $DualWindows$ and $Yack1$ are all selected from Weizmann data set. We scale them to $256 \times 256$.\par
1)$Plane$. We select the following parameters for MFSC: $R = 60$, $t = 7$, $\sigma_I = 8$, $\sigma_x = 4$, $\sigma_c = 0.095$, $\alpha = 0.45$. The corresponding parameters of Ncut are: $r = 15$, $\sigma_c = 0.1$.\par
2)$Imgp1883$. We select the following parameters for MFSC: $R = 60$, $t = 5$, $\sigma_I = 8$, $\sigma_x = 4$, $\sigma_c = 0.095$, $\alpha = 0.45$. The corresponding parameters of Ncut are: $r = 15$, $\sigma_c = 0.1$.\par
3)$DualWindows$. We select the following parameters for MFSC: $R = 60$, $t = 8$, $\sigma_I = 8$, $\sigma_x = 4$, $\sigma_c = 0.095$, $\alpha = 0.45$. The corresponding parameters of Ncut are: $r = 15$, $\sigma_c = 0.1$.\par
4)$Yack1$. We select the following parameters for MFSC: $R = 60$, $t = 8$, $\sigma_I = 8$, $\sigma_x = 4$, $\sigma_c = 0.095$, $\alpha = 0.45$. The corresponding parameters of Ncut are: $r = 15$, $\sigma_c = 0.1$.\par

\begin{figure*}[!t]
\centering
    \subfigure[]{
        \includegraphics[width=1.5in,height=1.5in]{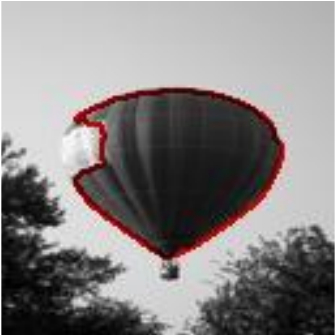}
    }
    \subfigure[]{
        \includegraphics[width=1.5in,height=1.5in]{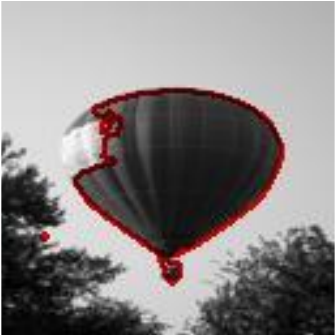}
    }
    \subfigure[]{
        \includegraphics[width=1.5in,height=1.5in]{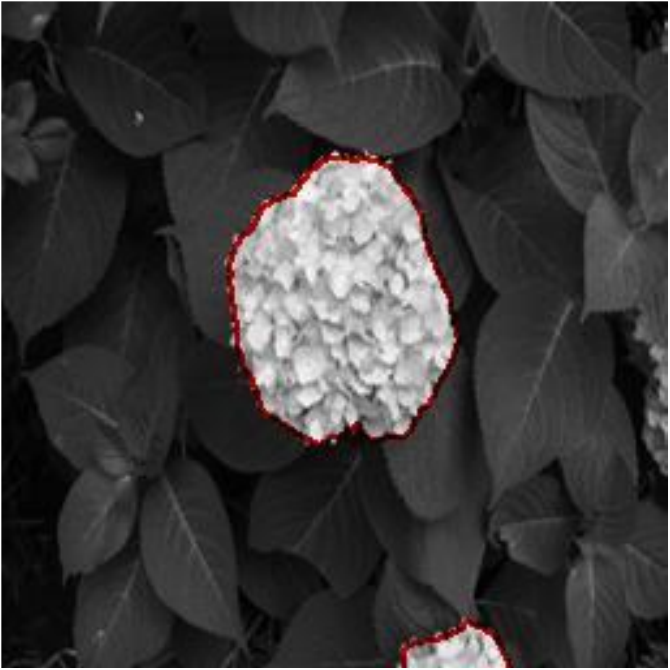}
    }
    \subfigure[]{
        \includegraphics[width=1.5in,height=1.5in]{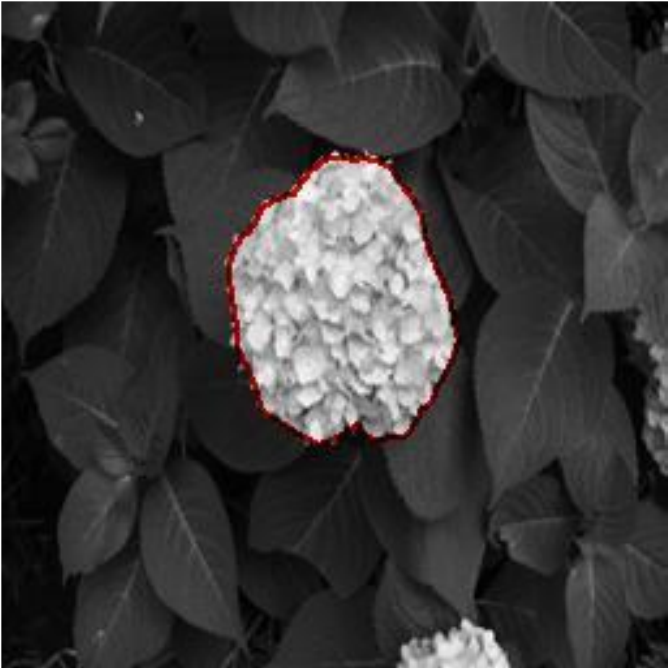}
    }
    \subfigure[]{
        \includegraphics[width=1.5in,height=1.5in]{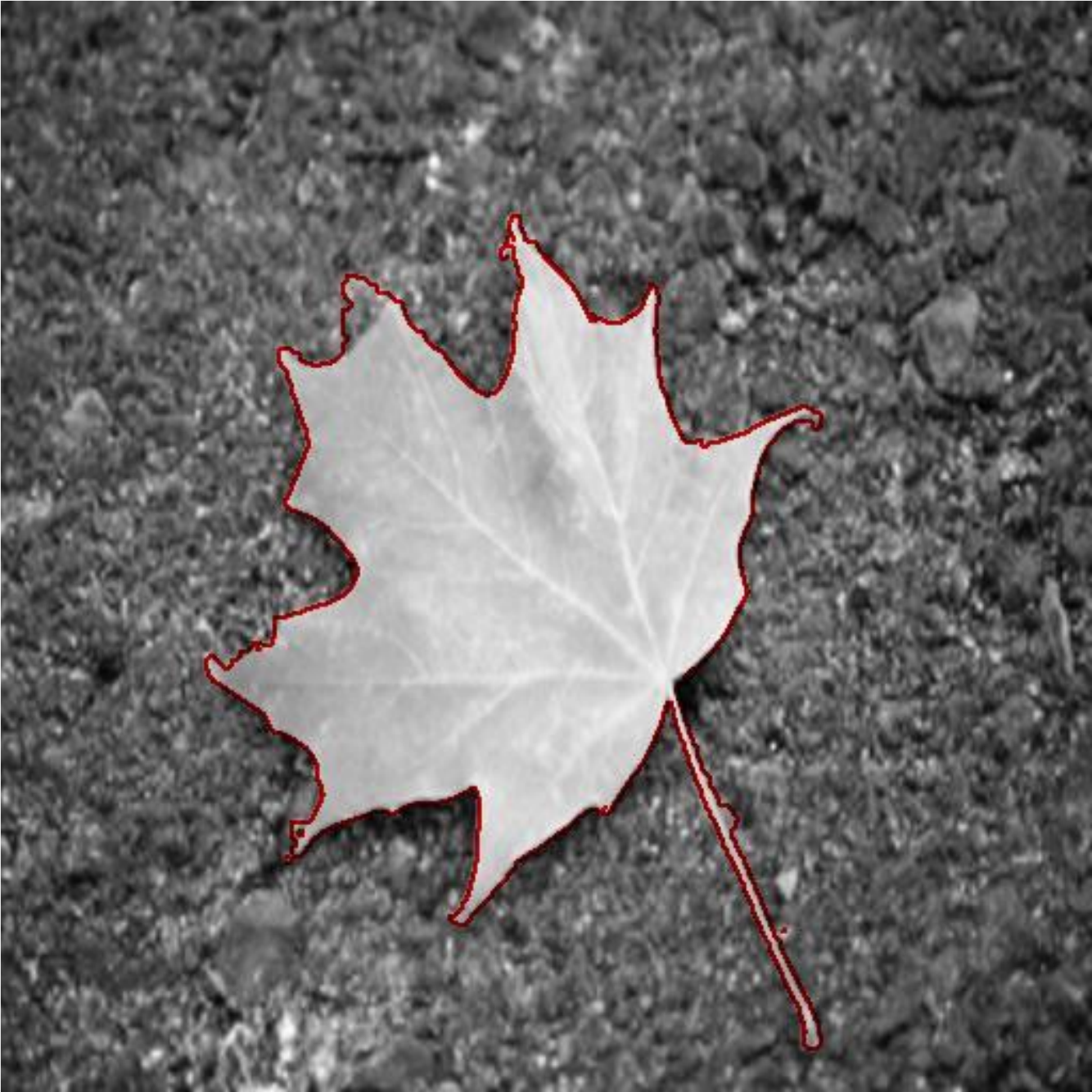}
    }
    \subfigure[]{
        \includegraphics[width=1.5in,height=1.5in]{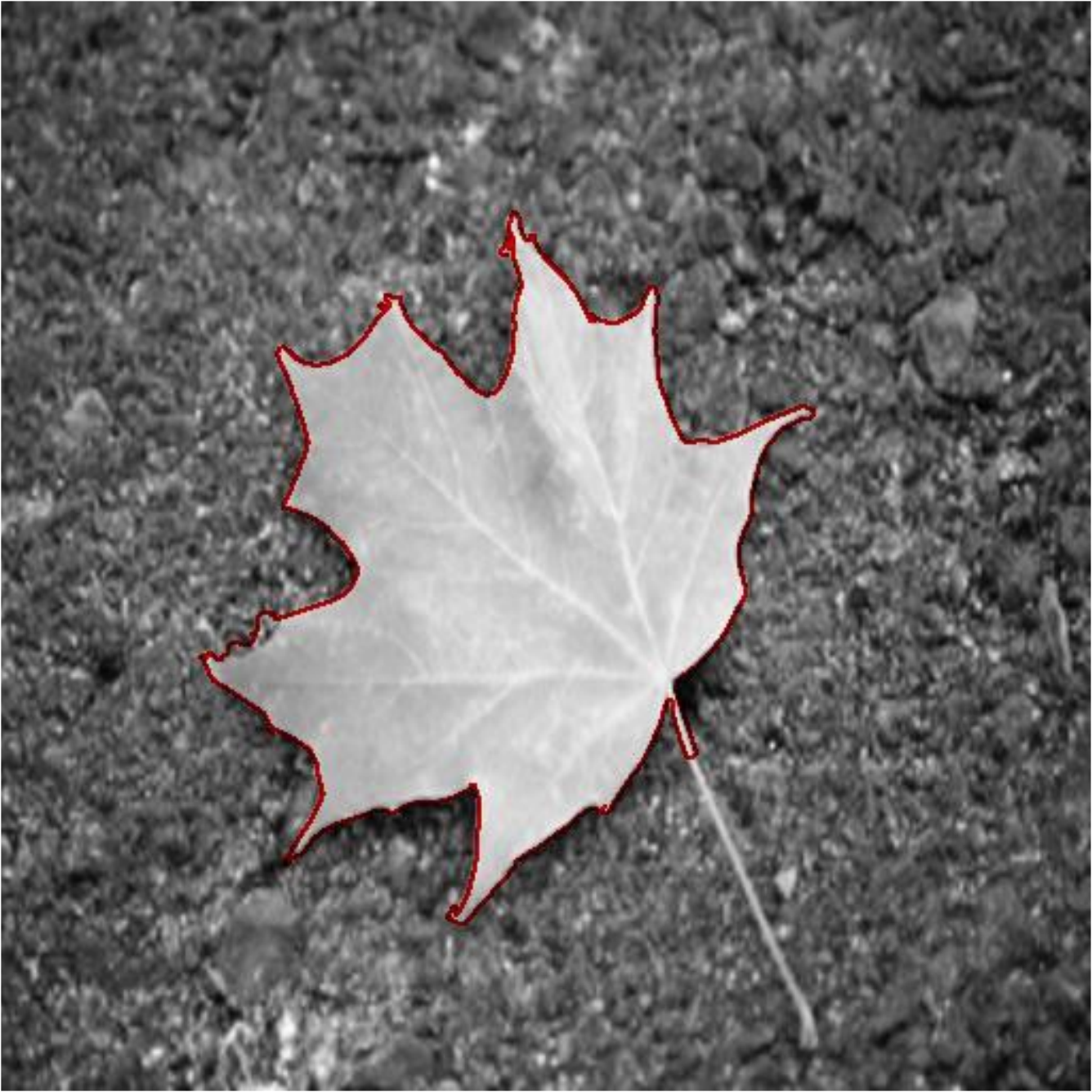}
    }
    \subfigure[]{
        \includegraphics[width=1.5in,height=1.5in]{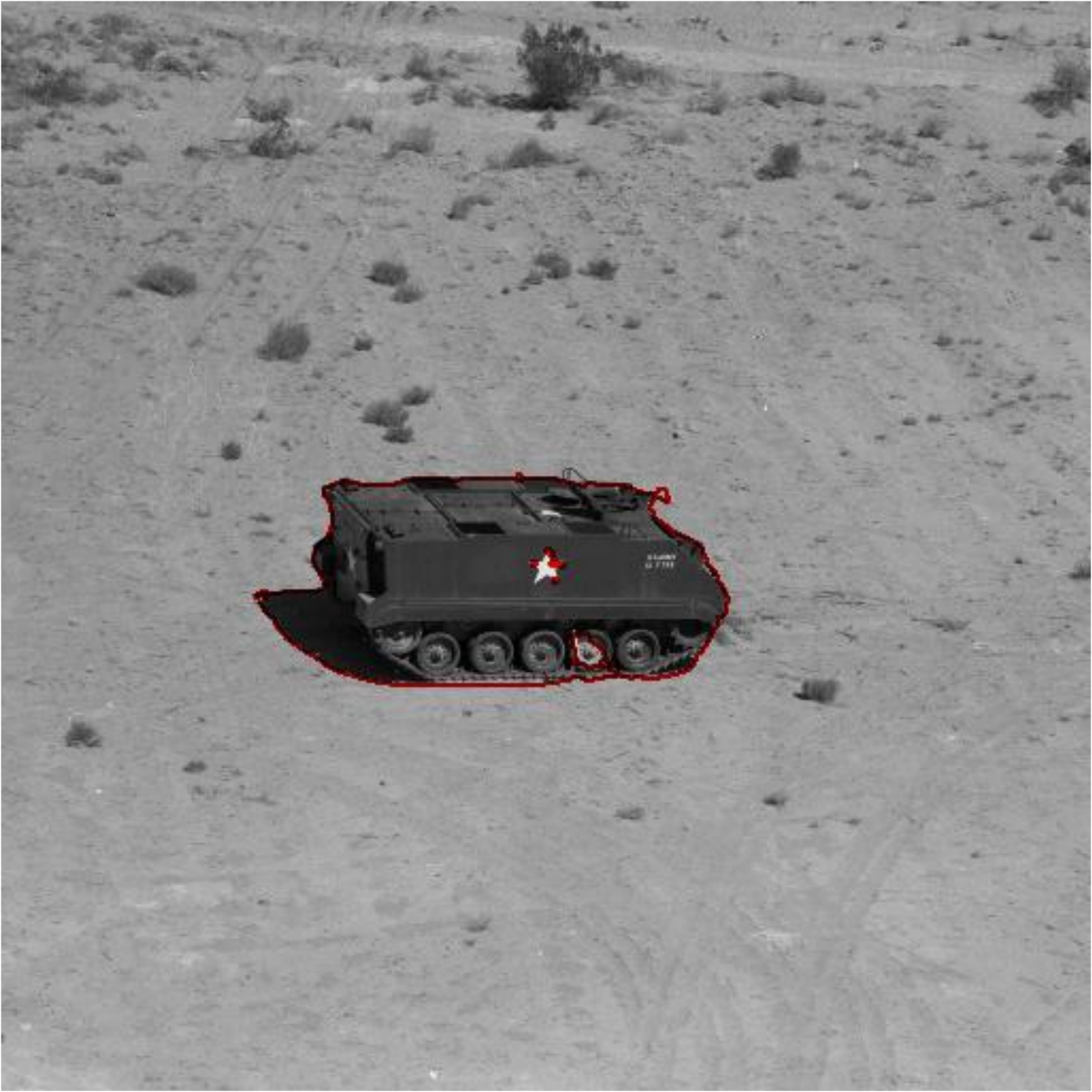}
    }
    \subfigure[]{
        \includegraphics[width=1.5in,height=1.5in]{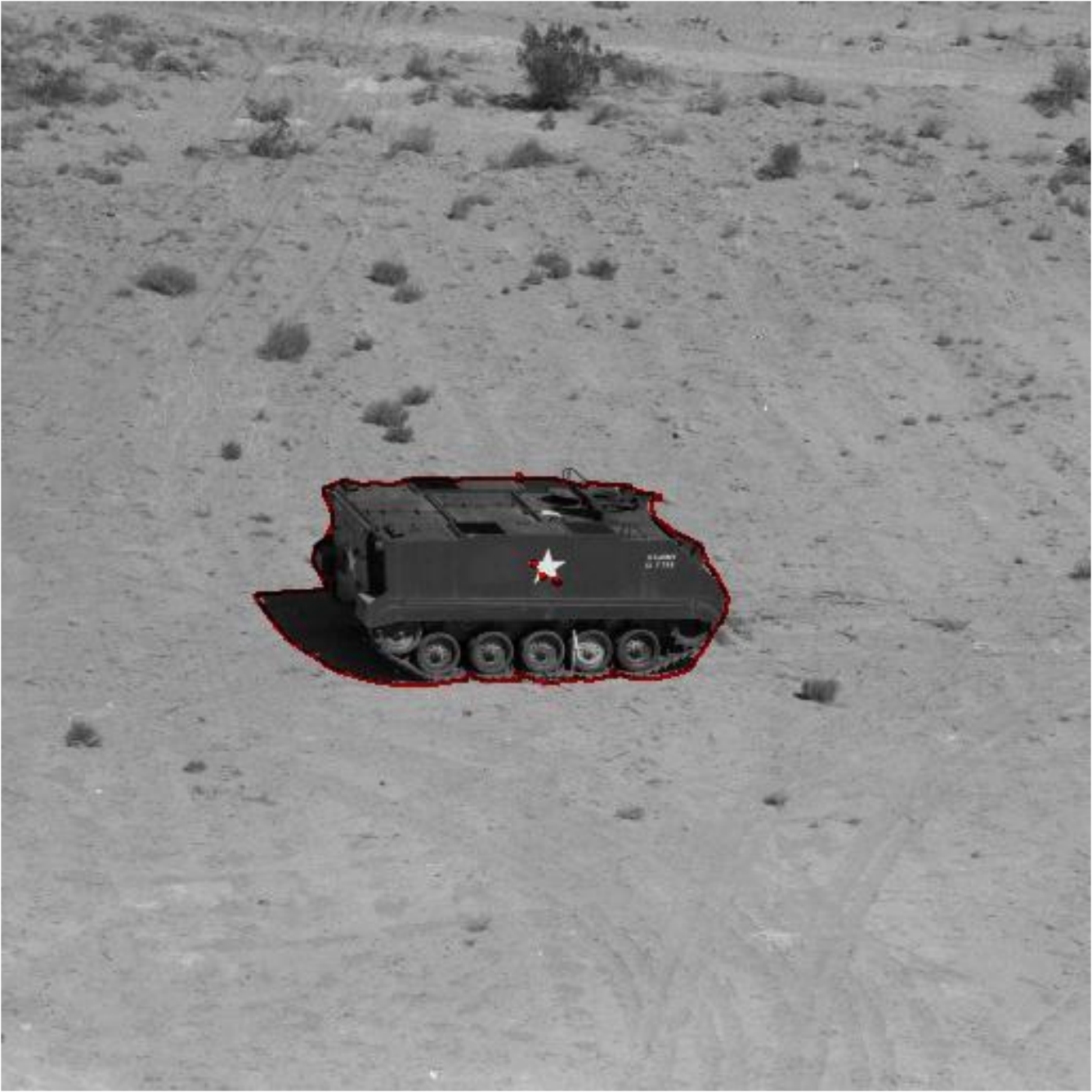}
        }
    \caption{The performances of MFSC and Ncut on four single-object images displayed in \figurename \ref{fig:four Image}. (a) (c) (e) (g): the results of MFSC; (b) (d) (f) (h): the results of Ncut.}
    \label{fig:four image results}
\end{figure*}

\begin{table*}[!t]
\centering
\renewcommand{\arraystretch}{1.3}
\caption{Segmentation Accuracy of the Four Images Displayed in \figurename \ref{fig:four Image}}
\label{table:performance in four images}
\centering
\begin{tabular}{|c|c|c|c|c|c|c|c|c|c|c|c|c|}
\hline
\multirow{2}*{\diagbox{\bfseries Method}{\bfseries Image name}} & \multicolumn{3}{c|}{\bfseries HotAirBalloon} &  \multicolumn{3}{c|}{\bfseries Nitpix} &  \multicolumn{3}{c|}{\bfseries Leafpav72} & \multicolumn{3}{c|}{\bfseries Tank}\\
\cline{2-13}
                 & \bfseries ACC & \bfseries DICE & \bfseries RI & \bfseries ACC & \bfseries DICE & \bfseries RI & \bfseries ACC & \bfseries DICE & \bfseries RI & \bfseries ACC & \bfseries DICE & \bfseries RI\\
\hline
\bfseries MFSC & 0.97 & 0.94 & 0.94 & 0.99& 0.96 & 0.98 & 0.99 & 0.97 & 0.98 & 0.99 & 0.93 & 0.98\\

\hline
\bfseries Ncut & 0.97 & 0.93 & 0.94 & 0.99 & 0.94 & 0.97 & 0.99 & 0.97 & 0.98 & 0.99 & 0.94 & 0.98\\
\hline

\end{tabular}
\end{table*}

\begin{table*}[!t]
\centering
\renewcommand{\arraystretch}{1.3}
\caption{Segmentation Accuracy of the Four Images Displayed in \figurename \ref{fig:object image}}
\label{table:performance in four 2 obj images}
\centering
\begin{tabular}{|c|c|c|c|c|c|c|c|c|c|c|c|c|}
\hline
\multirow{2}*{\diagbox{\bfseries Method}{\bfseries Image name}} & \multicolumn{3}{c|}{\bfseries Plane} &  \multicolumn{3}{c|}{\bfseries Imgp1883} &  \multicolumn{3}{c|}{\bfseries DualWindows} & \multicolumn{3}{c|}{\bfseries Yack1}\\
\cline{2-13}
                 & \bfseries ACC & \bfseries DICE & \bfseries RI & \bfseries ACC & \bfseries DICE & \bfseries RI & \bfseries ACC & \bfseries DICE & \bfseries RI & \bfseries ACC & \bfseries DICE & \bfseries RI\\
\hline
\bfseries MFSC & 0.96 & 0.85 & 0.90 & 0.98& 0.78 & 0.96 & 0.96 & 0.92 & 0.92 & 0.95 & 0.91 & 0.96\\

\hline
\bfseries Ncut & 0.95 & 0.83 & 0.91 & 0.98 & 0.79 & 0.97 & 0.96 & 0.94 & 0.87 & 0.98 & 0.94 & 0.93\\
\hline

\end{tabular}
\end{table*}


\begin{figure*}[!t]
\centering
    \subfigure[]{
        \includegraphics[width=1.5in,height=1.5in]{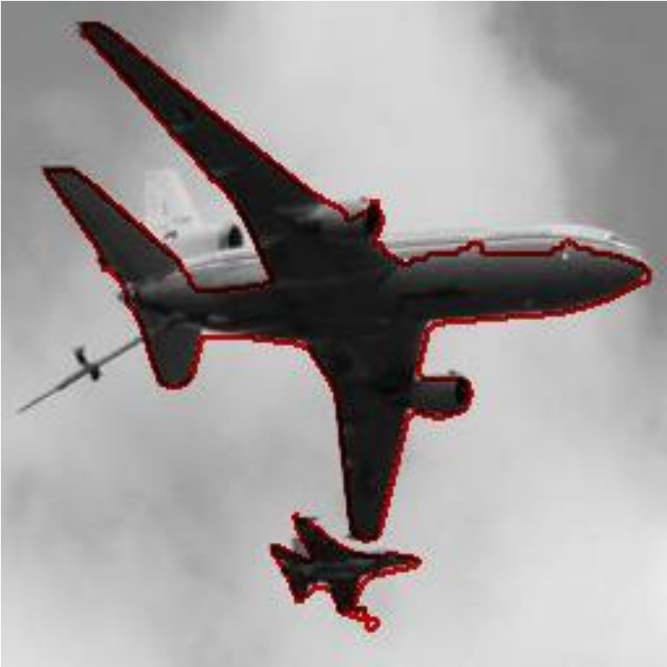}
    }
    \subfigure[]{
        \includegraphics[width=1.5in,height=1.5in]{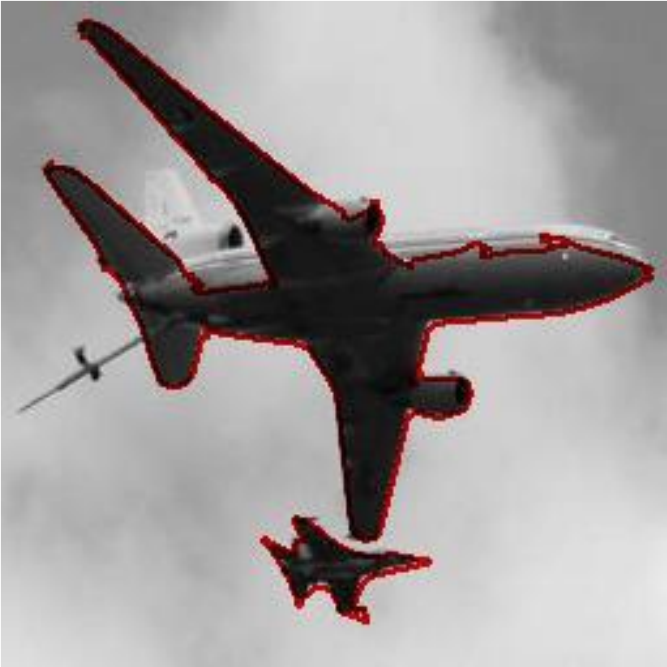}
    }
    \subfigure[]{
        \includegraphics[width=1.5in,height=1.5in]{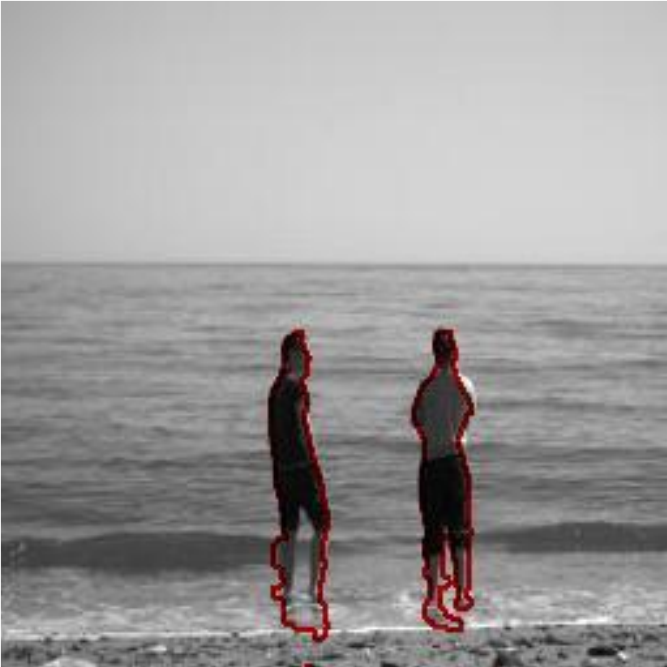}
    }
    \subfigure[]{
        \includegraphics[width=1.5in,height=1.5in]{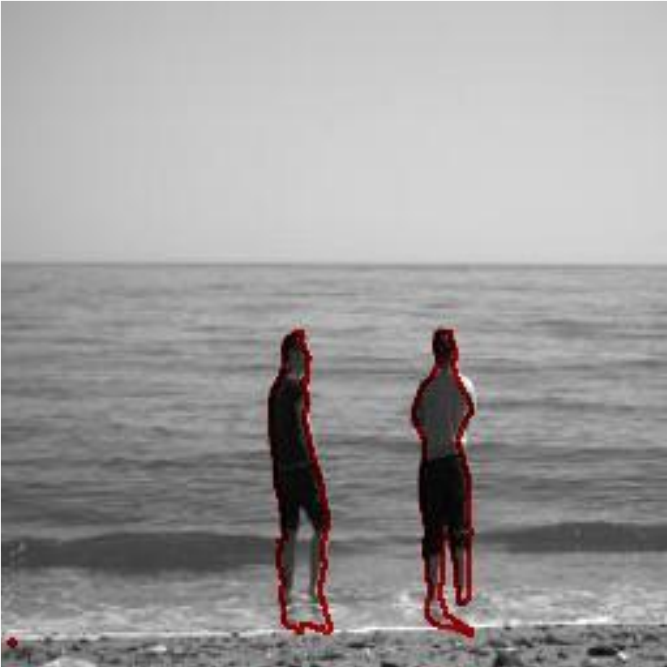}
    }
    \subfigure[]{
        \includegraphics[width=1.5in,height=1.5in]{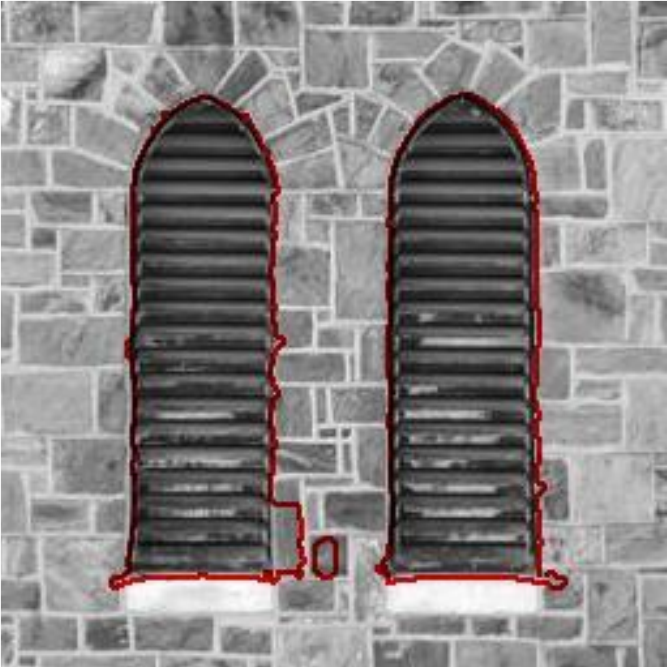}
    }
    \subfigure[]{
        \includegraphics[width=1.5in,height=1.5in]{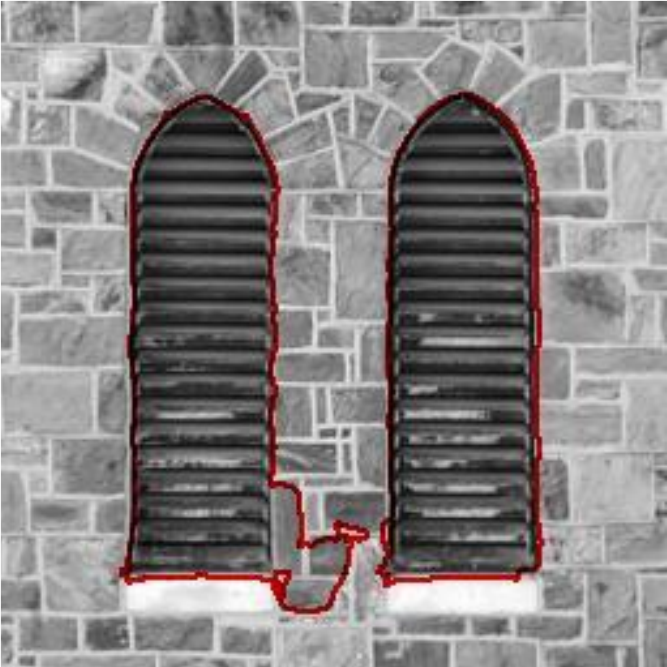}
        }
    \subfigure[]{
        \includegraphics[width=1.5in,height=1.5in]{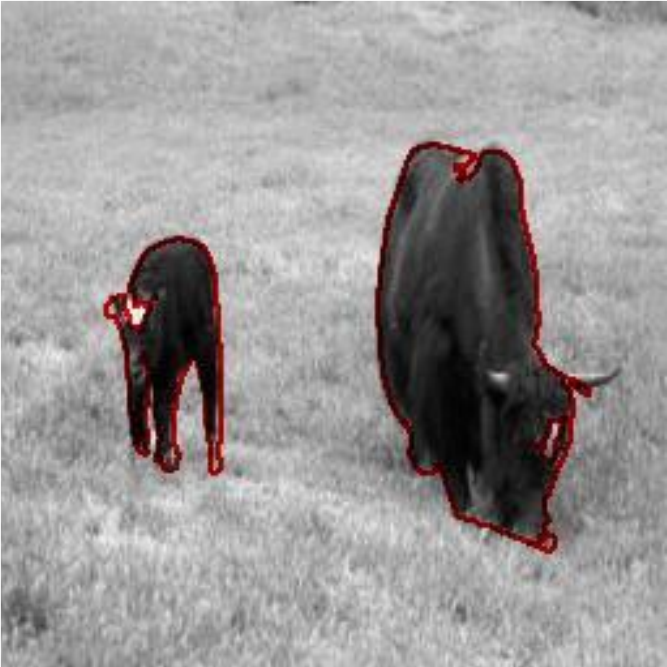}
    }
    \subfigure[]{
        \includegraphics[width=1.5in,height=1.5in]{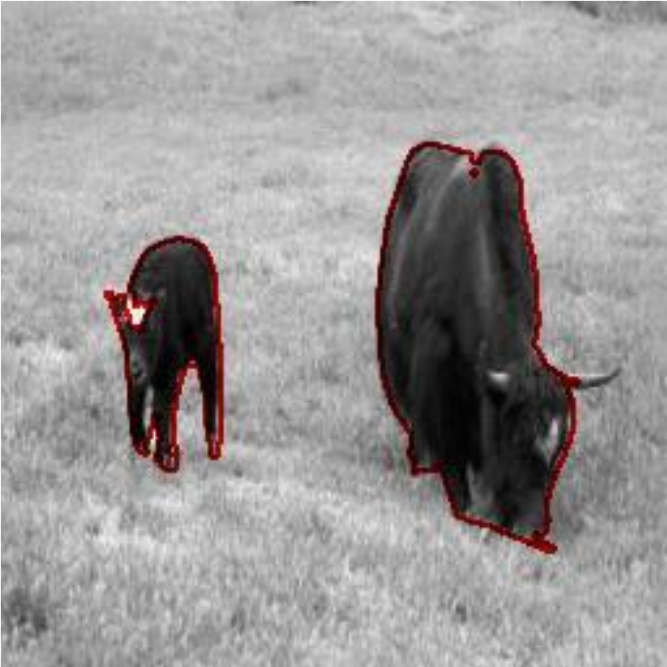}
    }

    \caption{The performances of MFSC and Ncut on four two-object images displayed in \figurename \ref{fig:object image}. (a) (c) (e) (g): the results of MFSC, (b) (d) (f) (h): the results of Ncut.}
    \label{fig:four Two object image results}
\end{figure*}

\subsection{Evaluation Metric}
We evaluate the segmentation quality by Accuracy (ACC) \cite{ACC}, Rand index (RI) \cite{RI} and Dice coefficient (Dice) \cite{Dice}. Given pixel $v_i$, let $o_i$, $s_i$, $F$, $B$ be its resultant segmentation label, ground-truth label, foreground and background, respectively.\par
$\bm {ACC}$: Define
\begin{equation}
\label{equ:acc}
ACC=\frac{\sum_{i=1}^n \delta (s_i\textrm{,}map(v_i\textrm{,}o_i))}{n}\textrm{,}
\end{equation}
where $\delta (a\textrm{,} b)$ denotes the delta function that returns $1$ if $a=b$ and $0$ otherwise; $map(v_i\textrm{,}o_i)$ is the best mapping function for permuting the cluster labels to match the ground-truth labels. The larger the ACC is, the better the segmentation performance is.\par
$\bm {RI}$: Define

\begin{equation}
\label{equ:RI}
RI=\frac{N+2T-P-Q}{N}\textrm{,}
\end{equation}

where
\begin{equation*}
\begin{aligned}
P&=\sum_{i\in \{F\textrm{,}B\}}\left(\begin{array}{c}
    \sum_{i\in \{F\textrm{,}B\}} m_{ij} \\
    2
  \end{array}\right)\textrm{,}\\
Q&=\sum_{j\in \{F\textrm{,}B\}}\left(\begin{array}{c}
    \sum_{j\in \{F\textrm{,}B\}} m_{ij} \\
    2
  \end{array}\right)\textrm{,}\\
N&=\left(\begin{array}{c}
    \sum_{i\textrm{,}j\in \{F\textrm{,}B\}} m_{ij} \\
    2
  \end{array}\right)\textrm{,}\\
T&=\frac{1}{2}[\sum_{i\textrm{,}j\in \{F\textrm{,}B\}} m_{ij}^2-\sum_{i\textrm{,}j\in \{F\textrm{,}B\}} m_{ij}]\textrm{,}
\end{aligned}
\end{equation*}
where $m_{ij}=|o_i\bigcap s_j|\ $, $i\textrm{,}j\in \{F\textrm{,}B\}$. The range of RI is in the interval of 0 and 1, where 0 is for absolute mismatch and 1 for equality
to the ground truth.\par
$\bm {Dice}$: Define
\begin{equation}
\begin{aligned}
Dice=\frac{|o_i\bigcap s_i|}{|o|+|s|}.
\end{aligned}
\end{equation}
Its range is  from 0 to 1 (1 for perfect match with ground truth).\par
 All of our experiments are conducted on a Windows 10 x64 computer with a 3.4 GHz Intel(R) Core(TM) i7-3770 CPU and 8 GB RAM, MATLAB.\par

\subsection{Parameter Analysis}

\begin{figure*}[!t]
\centering
    \subfigure[]{
        \includegraphics[width=1.5in,height=1.5in]{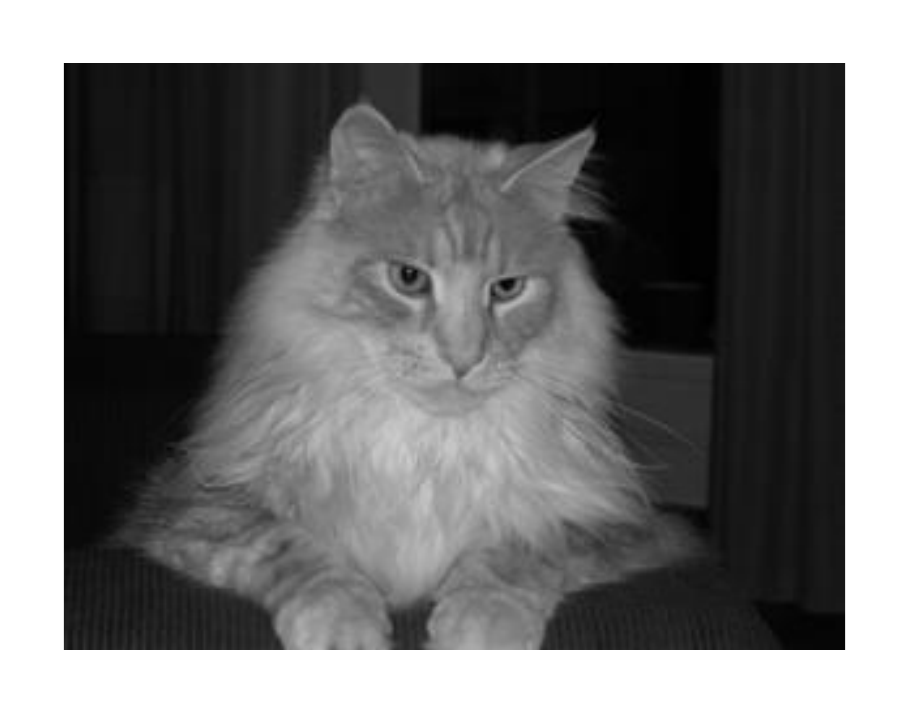}
    }
    \subfigure[]{
        \includegraphics[width=1.5in,height=1.5in]{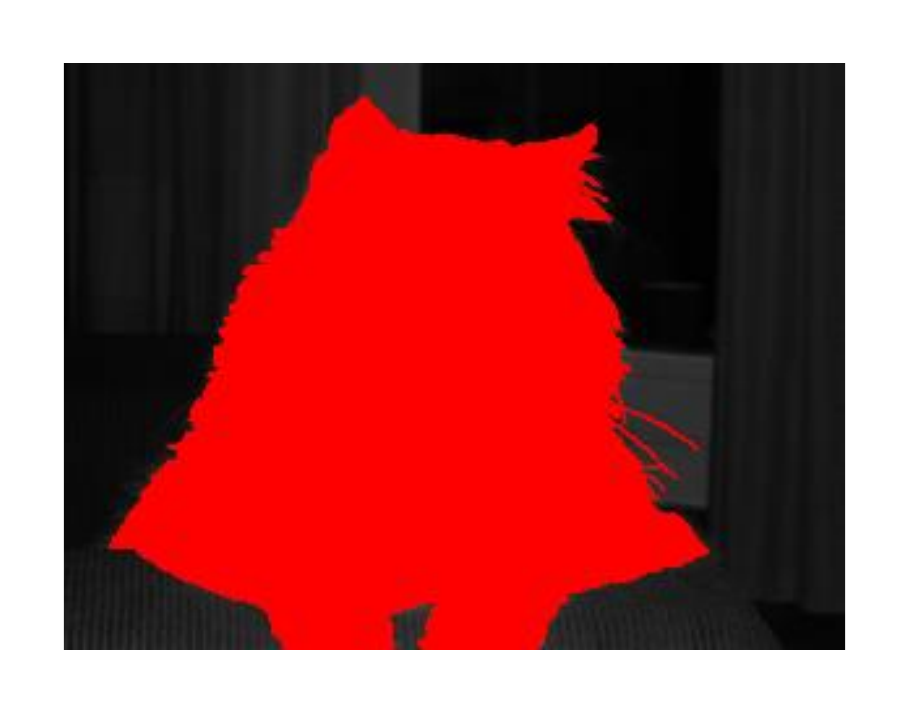}
    }
    \subfigure[]{
        \includegraphics[width=1.5in,height=1.5in]{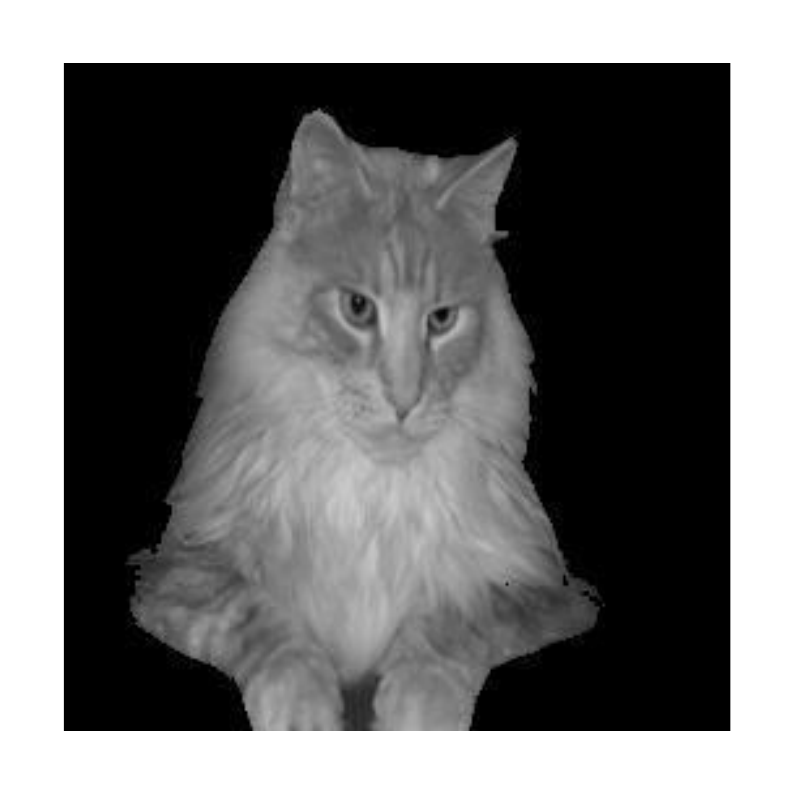}
    }
    \subfigure[]{
        \includegraphics[width=1.5in,height=1.5in]{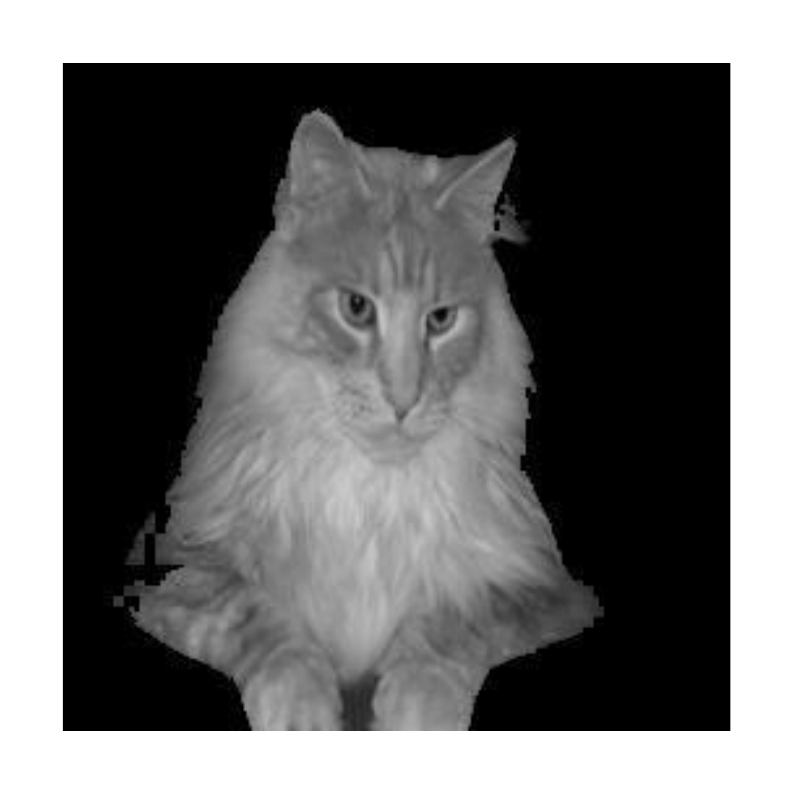}
    }
    \subfigure[]{
        \includegraphics[width=1.5in,height=1.5in]{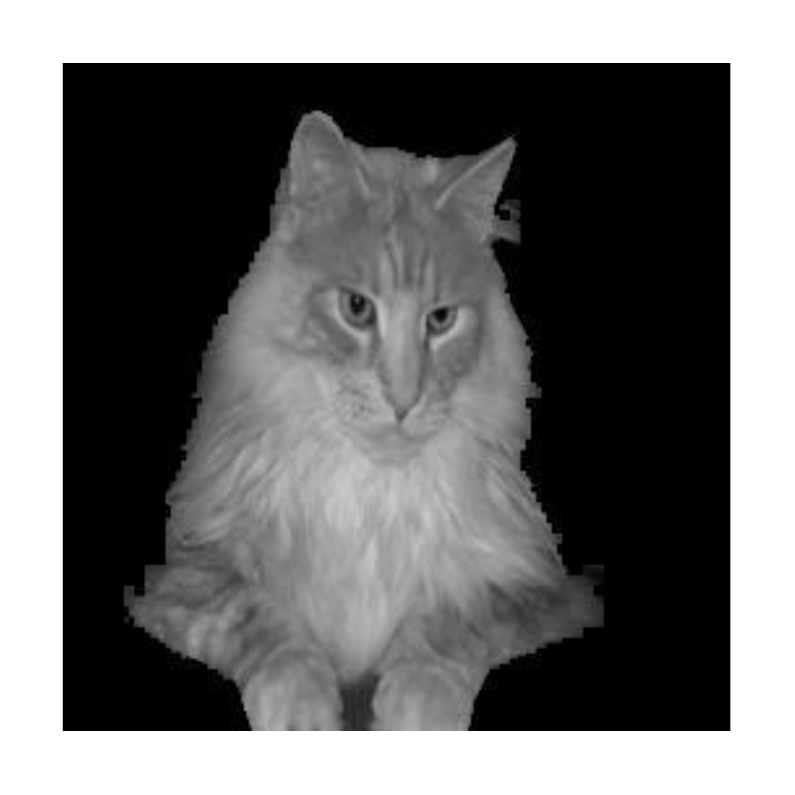}
    }
        \subfigure[]{
        \includegraphics[width=1.5in,height=1.5in]{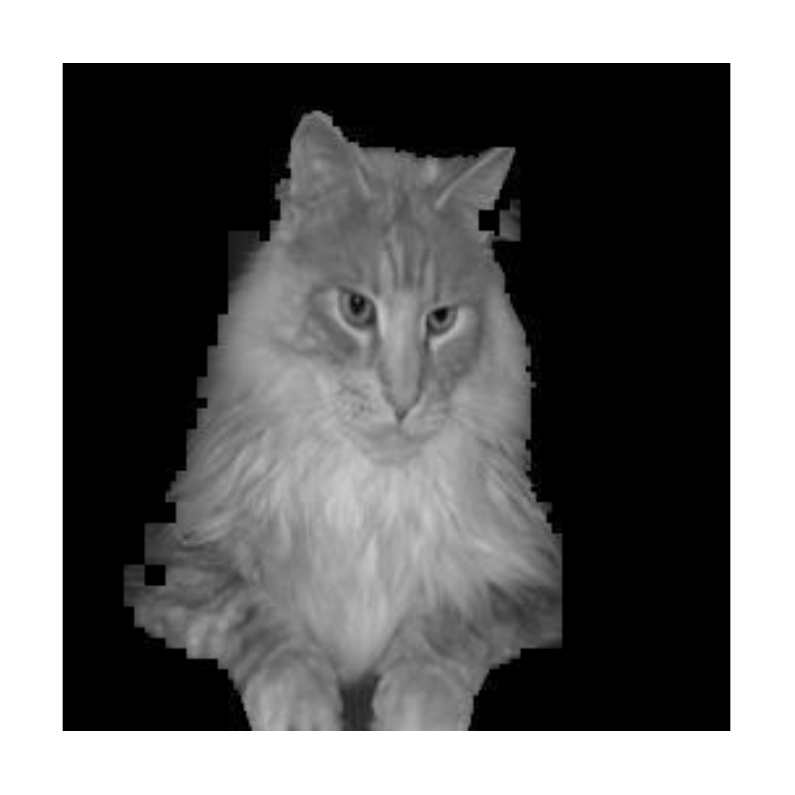}
    }
        \subfigure[]{
        \includegraphics[width=1.5in,height=1.5in]{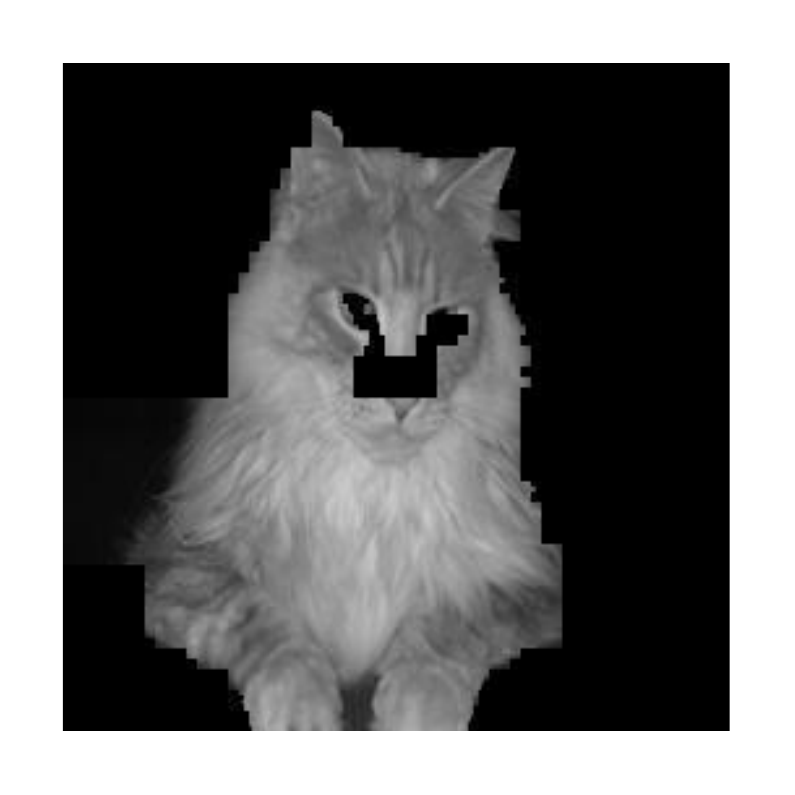}
    }
    \caption{The segmentation results of the image under different thresholds. (a) the original image ($256 \times 256$), (b) ground truth, (c)-(g): the segmentation results of MFSC under the threshold of 2, 6, 10, 15, 20 respectively.}
    \label{fig:cat}
\end{figure*}

\begin{figure*}[!t]

\centering
    \subfigure[]{
        \includegraphics[width=3in,height=2.25in]{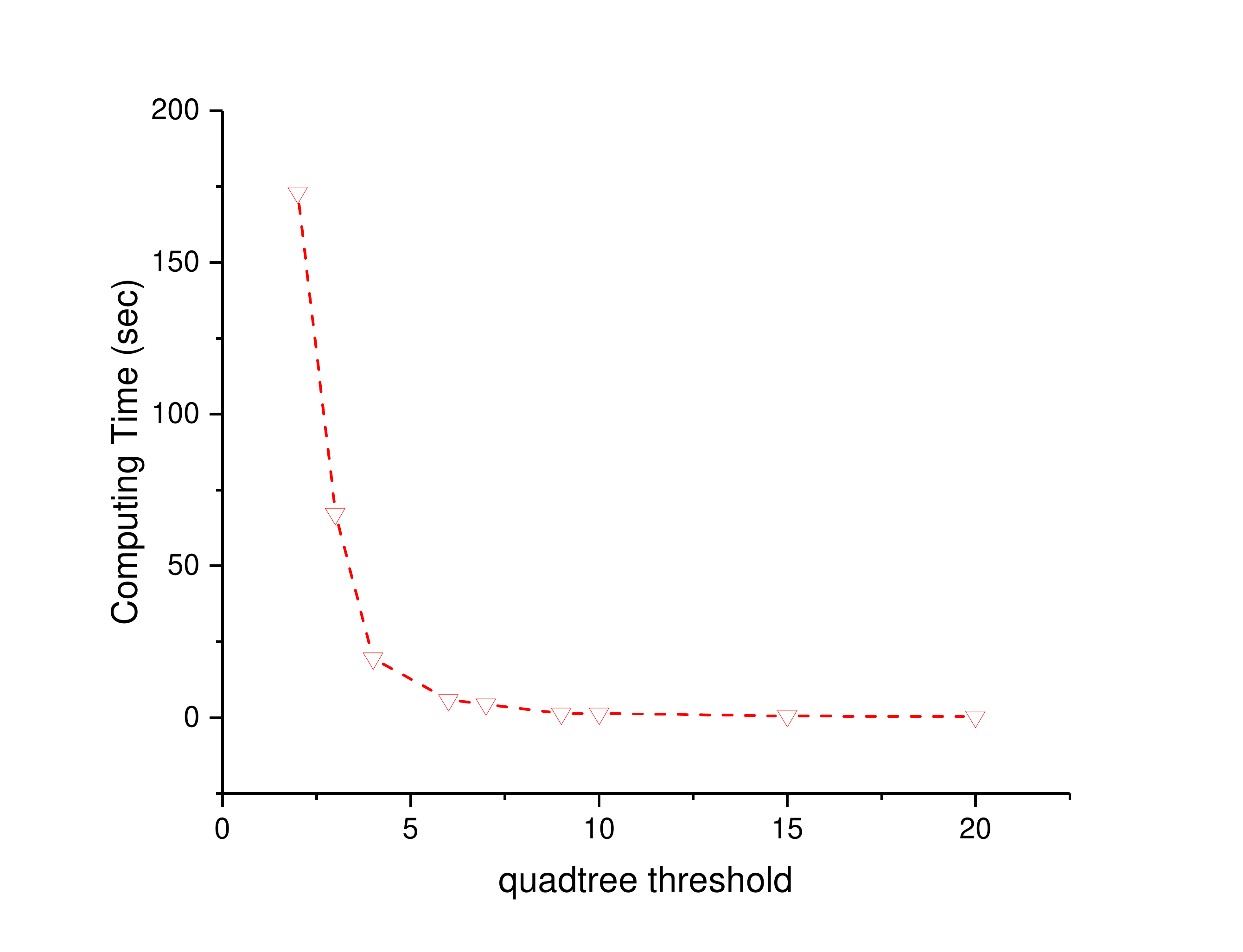}
    }
    \subfigure[]{
        \includegraphics[width=3in,height=2.25in]{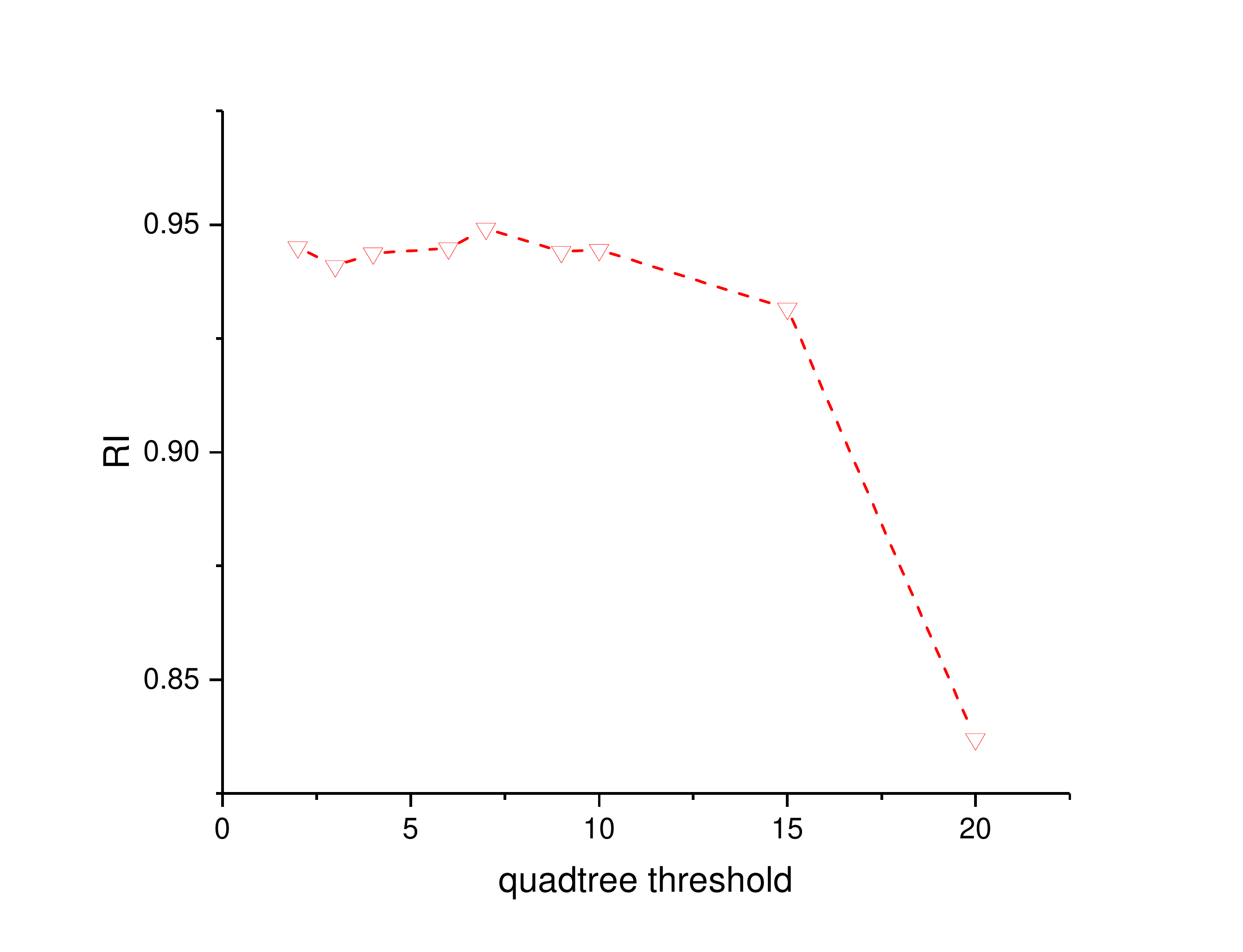}
    }
    \caption{The relationship between the threshold of quad-tree decomposition and (a) computing time and (b) segmentation performance of MFSC in \figurename \ref{fig:cat}.      }
    \label{fig:relation}
\end{figure*}

We study the relationship between the threshold of quad-tree decomposition and the performance of MFSC. We take an image from Weizmann data set as an example. \figurename \ref{fig:cat} shows the segmentation result of the image under different thresholds of quad-tree decomposition with MFSC. It can be observed that the larger the threshold is, the worse the result is, the shorter the computing time is. The relationship is presented more clearly via the parameters of computing time and RI in \figurename \ref{fig:relation}. However, it can also be observed that the segmentation performance is robust to the threshold when the threshold is smaller than a certain value. The reason behind is that under this circumstances, the quad-tree structure can show the complete texture/shape cues when the threshold is below some value. We empirically set the threshold for $128 \times 128$ images to be $10$, $256 \times 256$ $12$, $512 \times 512$ $16$, which is appropriate to most images used in our experiments.   \par

\begin{figure*}[!t]
\centering
    \subfigure[]{
        \includegraphics[width=1.5in,height=1.5in]{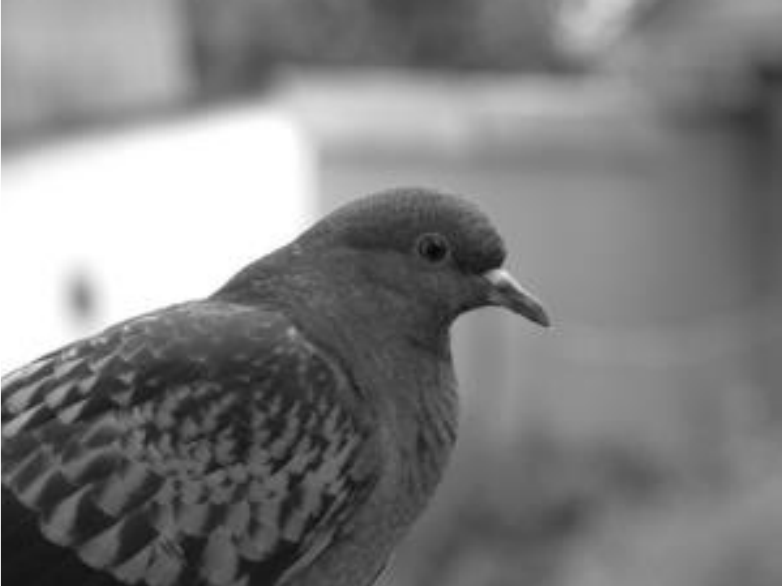}
        \label{fig:birdsource}
    }
    \subfigure[]{
        \includegraphics[width=1.5in,height=1.5in]{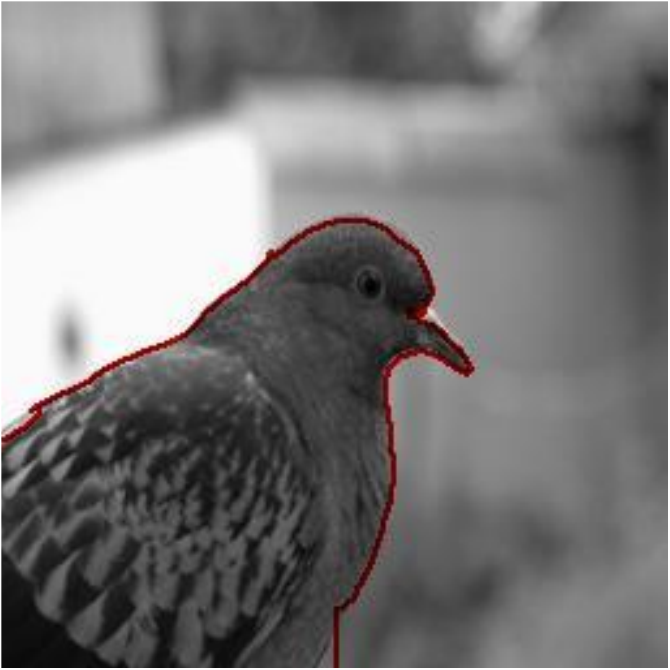}
    }
    \subfigure[]{
        \includegraphics[width=1.5in,height=1.5in]{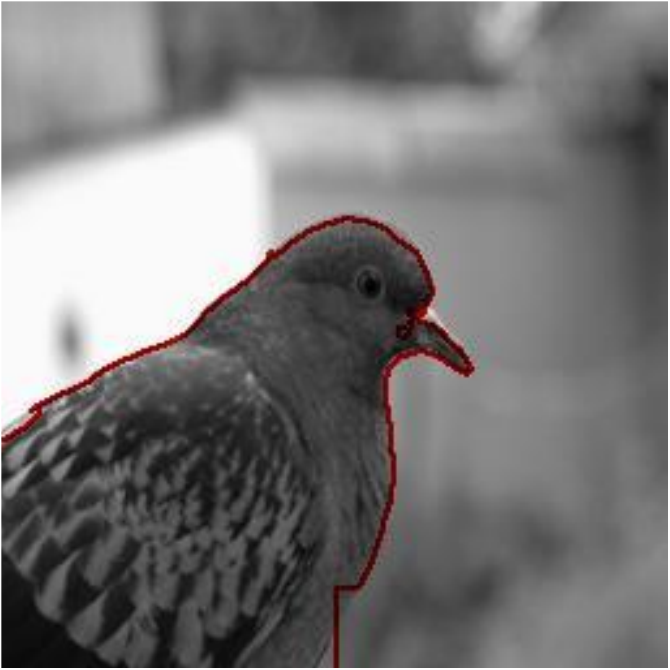}
    }
    \subfigure[]{
        \includegraphics[width=1.5in,height=1.5in]{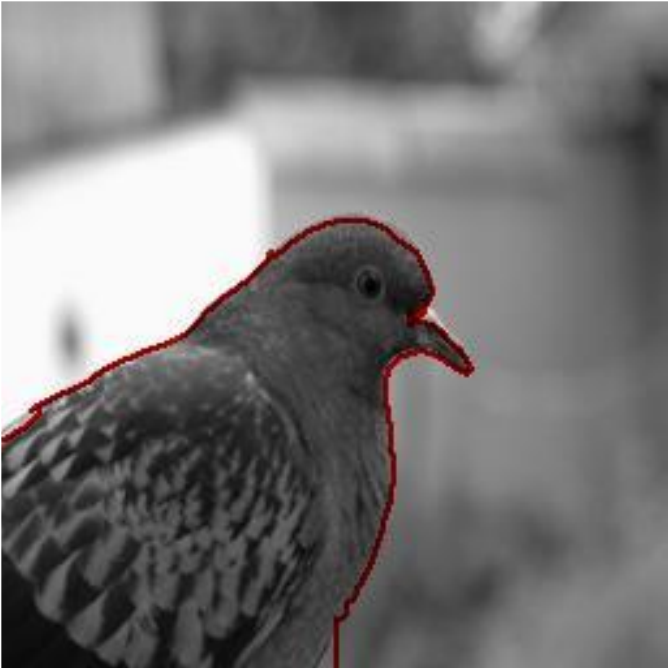}
    }
    \subfigure[]{
        \includegraphics[width=1.5in,height=1.5in]{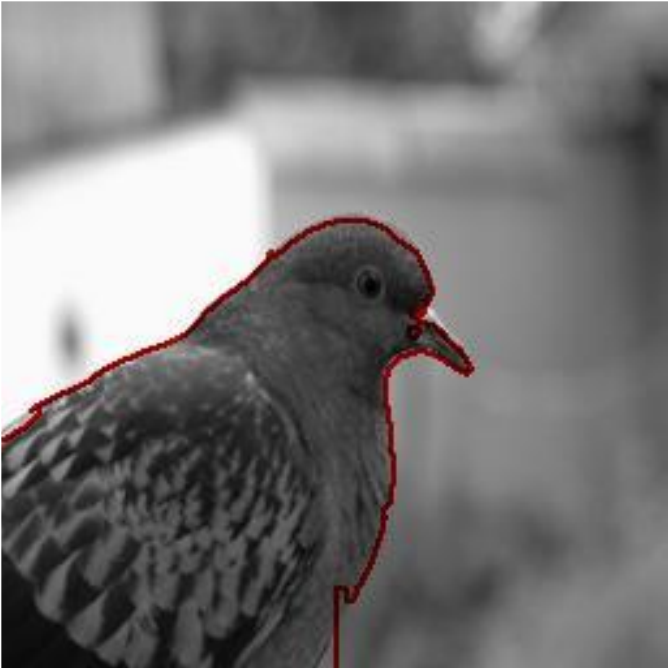}
    }
        \subfigure[]{
        \includegraphics[width=1.5in,height=1.5in]{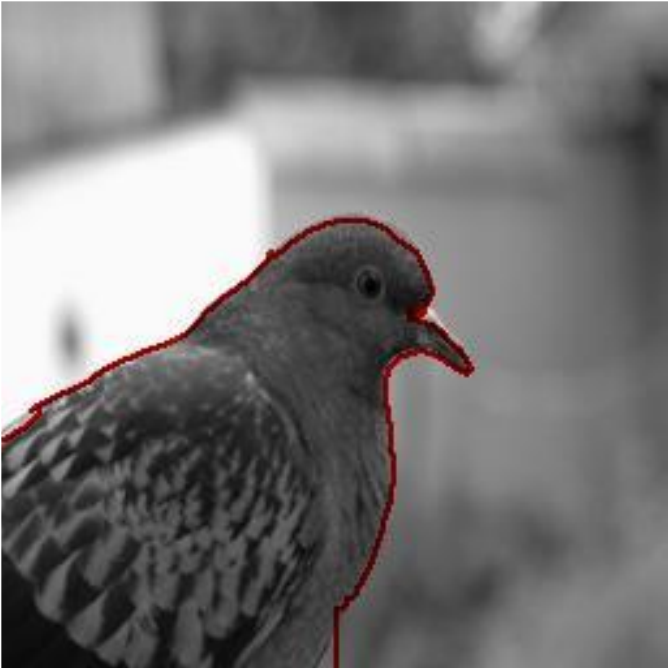}
    }
        \subfigure[]{
        \includegraphics[width=1.5in,height=1.5in]{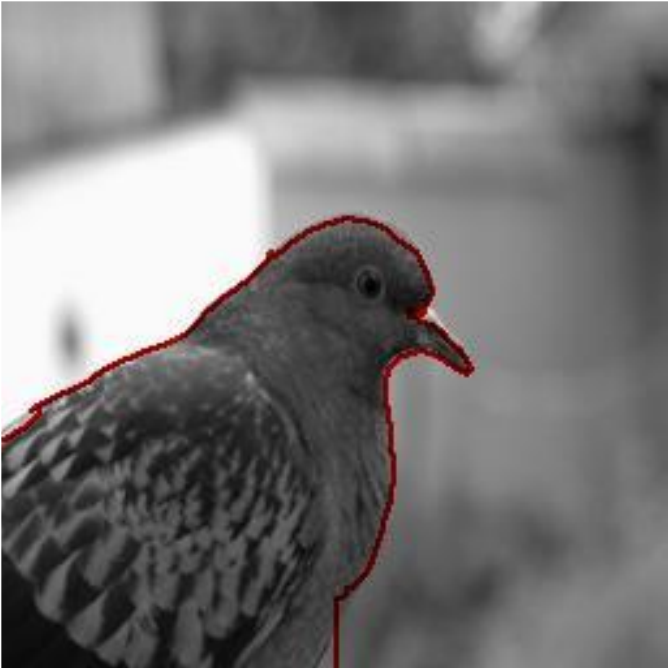}
    }
    \caption{The segmentation results of the image under different start levels. (a) the original image ($256 \times 256$), (b)-(g): the segmentation results of MFSC with the start level being 2, 3, 4, 5, 6, 7, respectively.}
    \label{fig:bird}
\end{figure*}

\begin{figure*}[!t]

\centering
    \subfigure[]{
        \includegraphics[width=3in,height=2.25in]{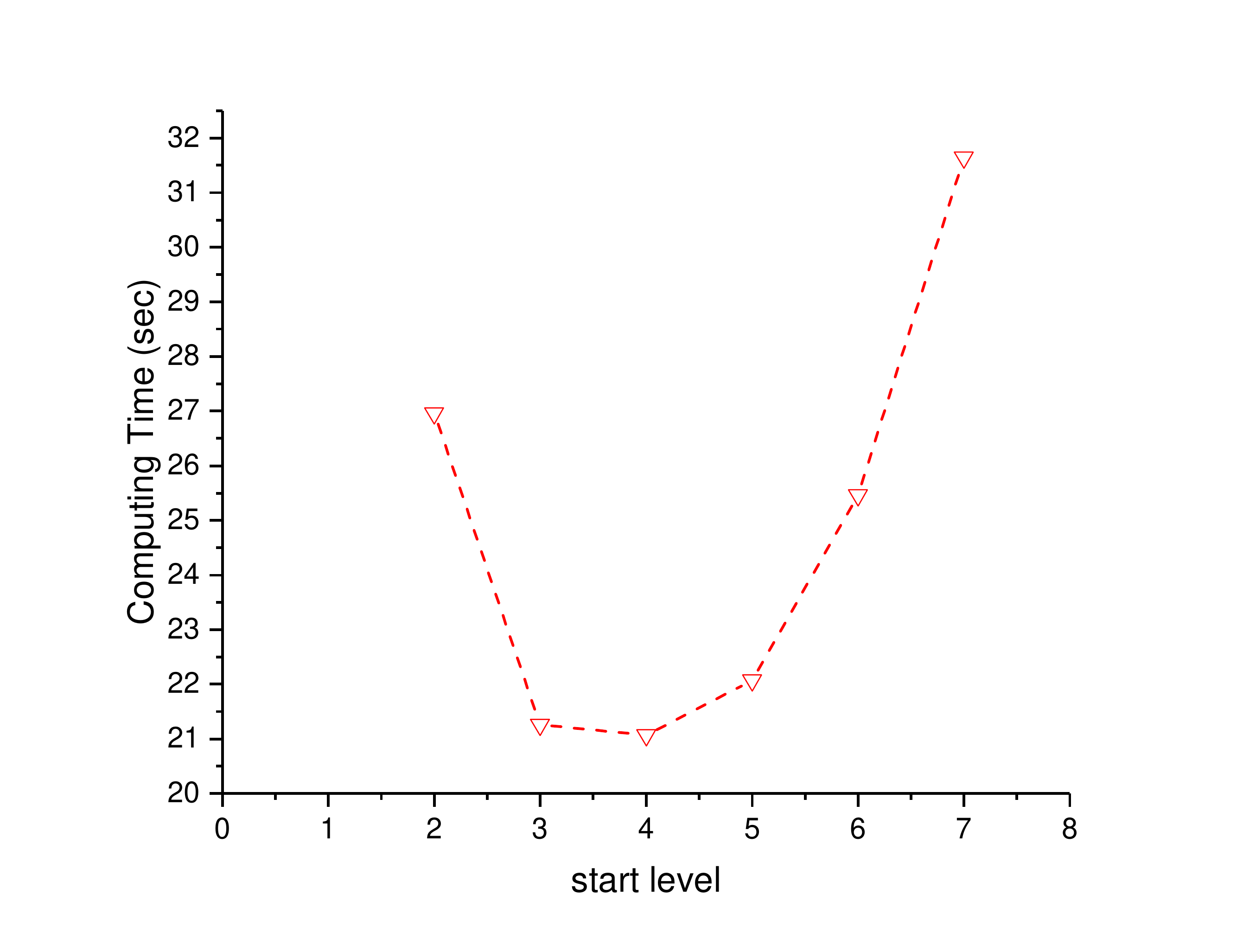}
    }
    \subfigure[]{
        \includegraphics[width=3in,height=2.25in]{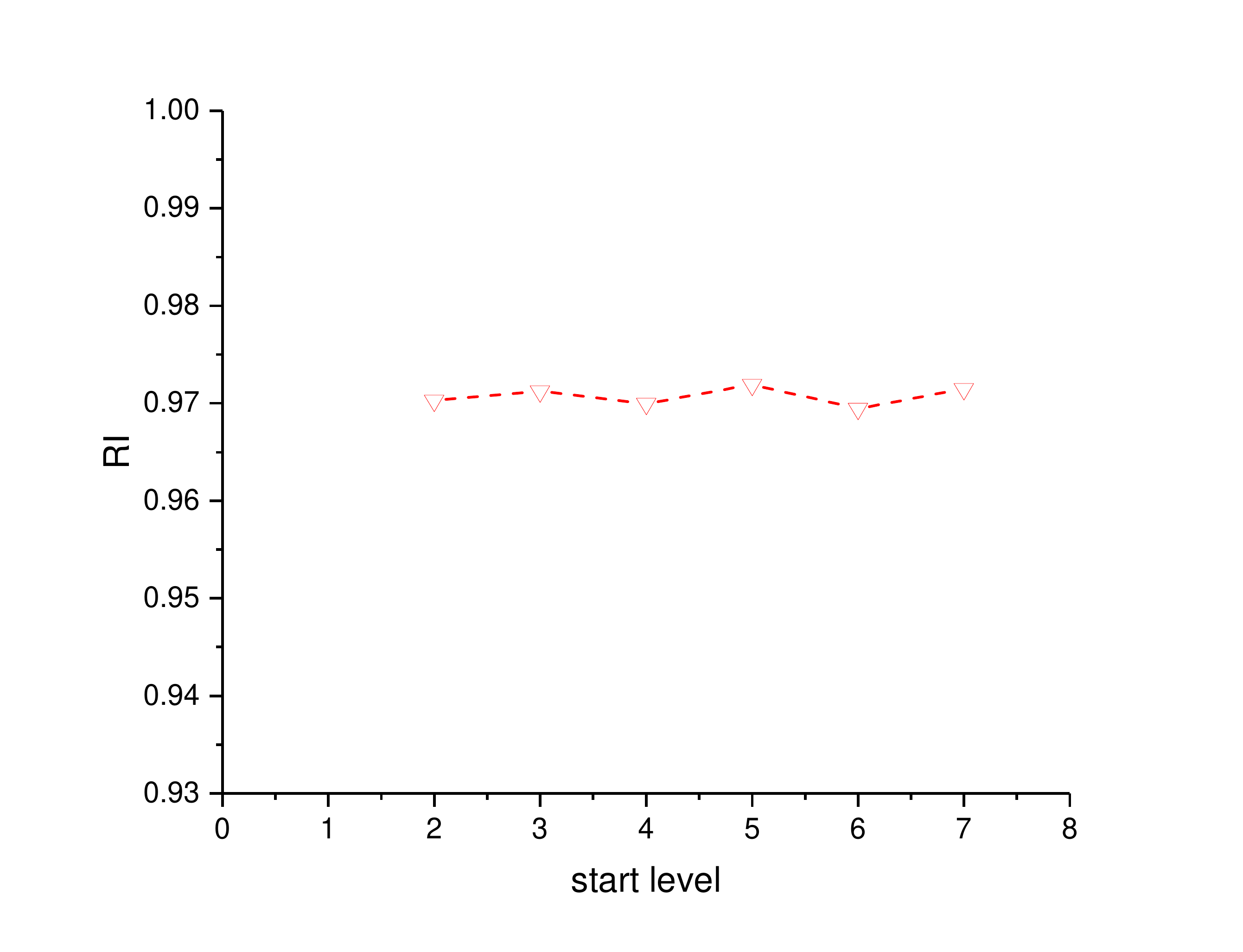}
    }
    \caption{The relationship between the start level and (a) computing time and (b) segmentation performance of MFSC in \figurename \ref{fig:bird}.      }
    \label{fig:relation of start level}
\end{figure*}

Next, we will explain how to choose the start level $l_{init}$. We select an image from Weizmann data set which is shown in \figurename \ref{fig:birdsource}. The parameters of MFSC are as follow: $R = 60$, $t = 5$, $\sigma_I = 8$, $\sigma_x = 4$, $\sigma_c = 0.15$, $\alpha = 0.45$. We resize this image to $256\times256$ and test MFSC with different start levels from second level to seventh level.  \figurename \ref{fig:bird} and \figurename \ref{fig:relation of start level} show that the segmentation result of MFSC is robust to the start level. Hence, we empirically choose the third or forth level as the start level.

\begin{figure}[!t]
\centering

    \includegraphics[width=4in,height=3in]{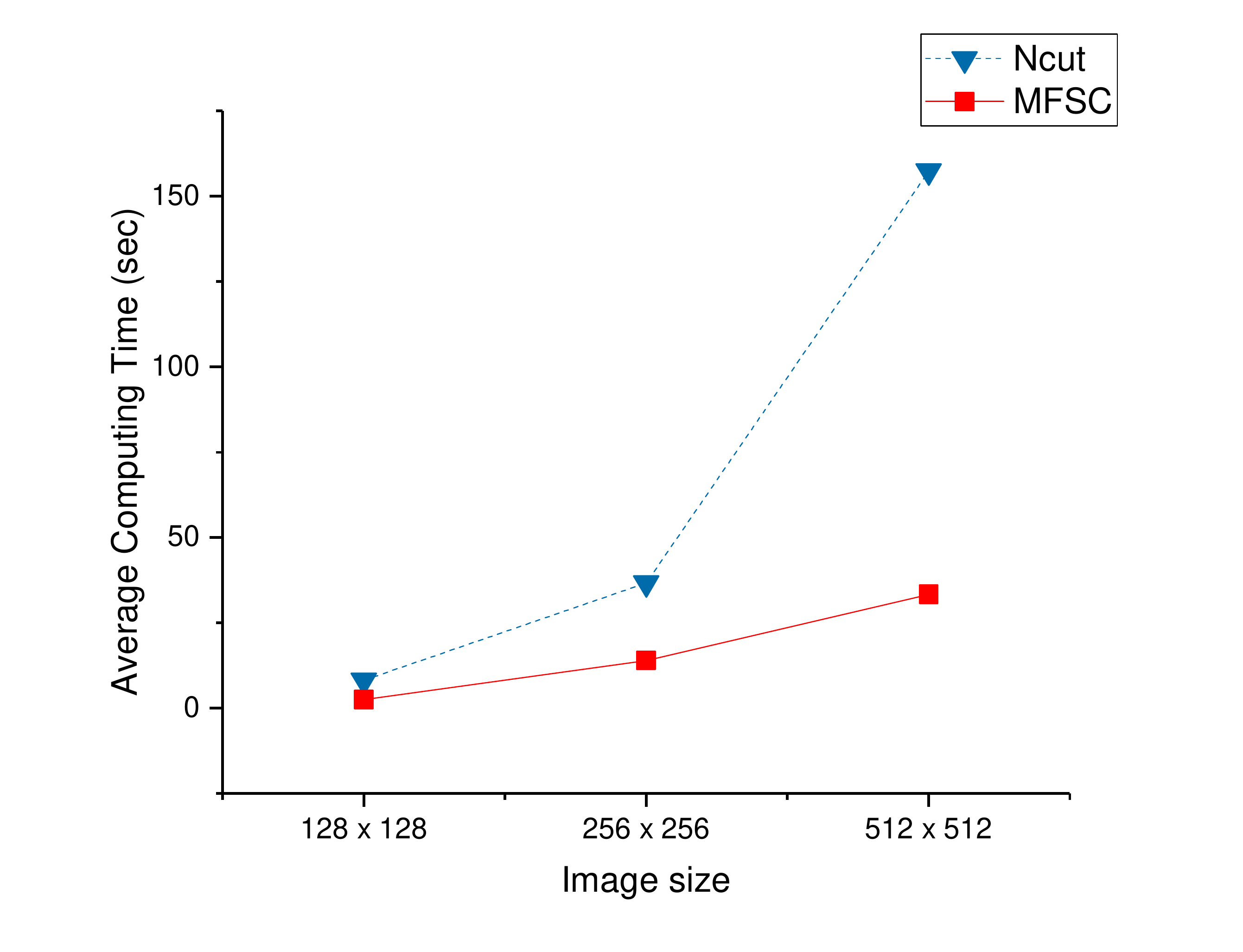}
    \caption{The average computing time of Ncut and MFSC on images of different sizes.}
    \label{fig:average time}
\end{figure}


\begin{figure*}[!t]
\centering
     \subfigure[]{
        \includegraphics[width=5in,height=2in]{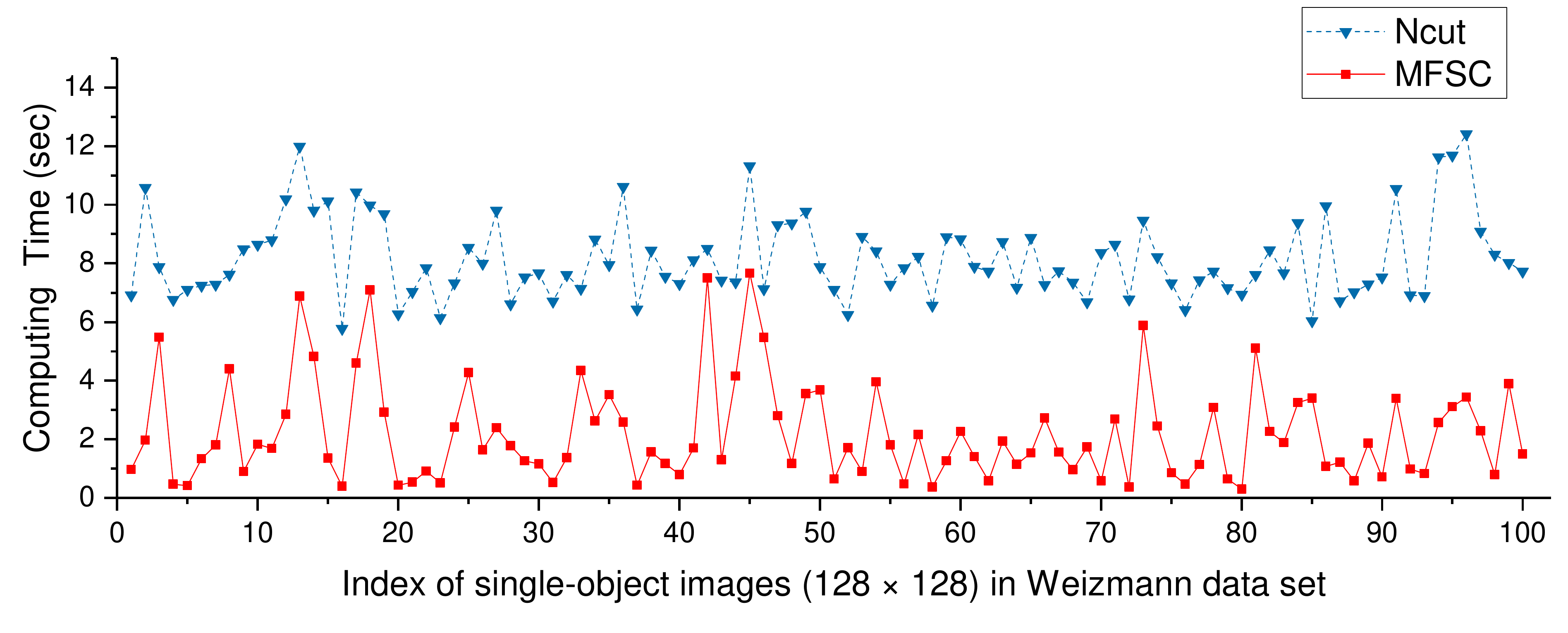}
        }
     \subfigure[]{
        \includegraphics[width=5in,height=2in]{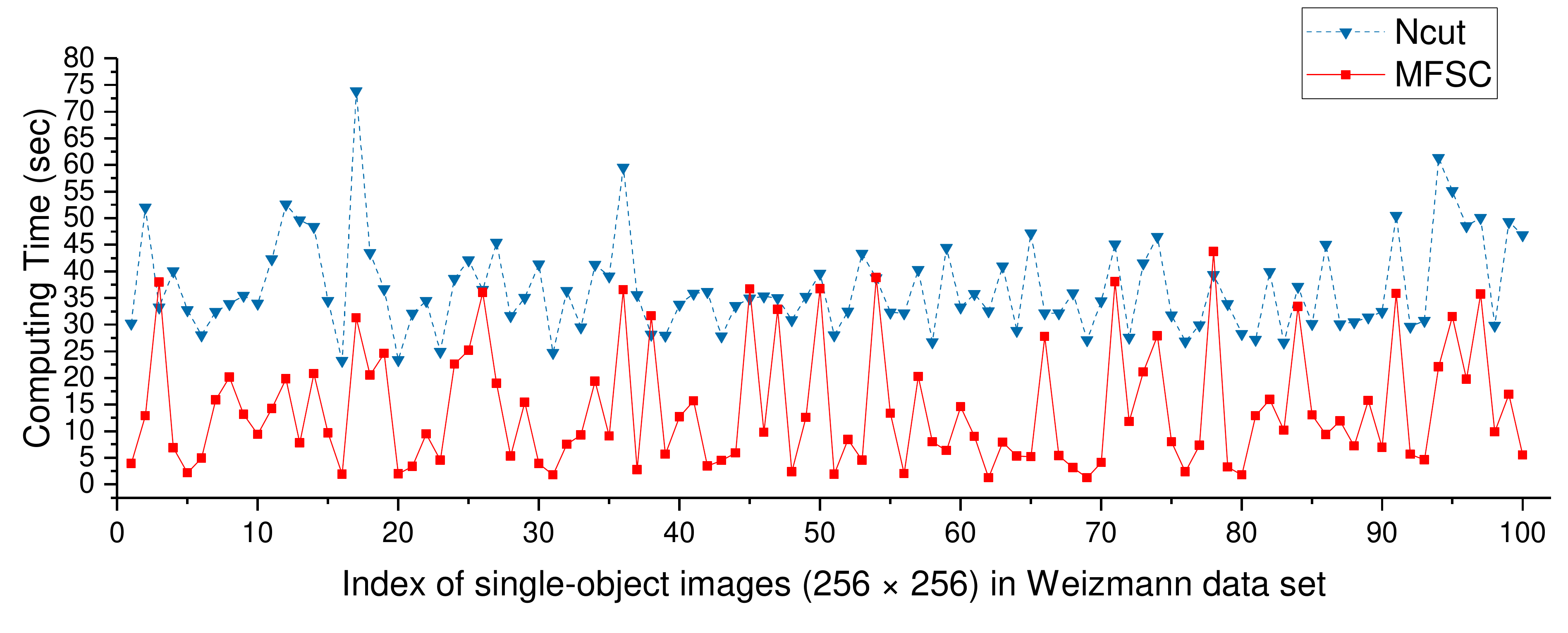}
        }
     \subfigure[]{
        \includegraphics[width=5in,height=2in]{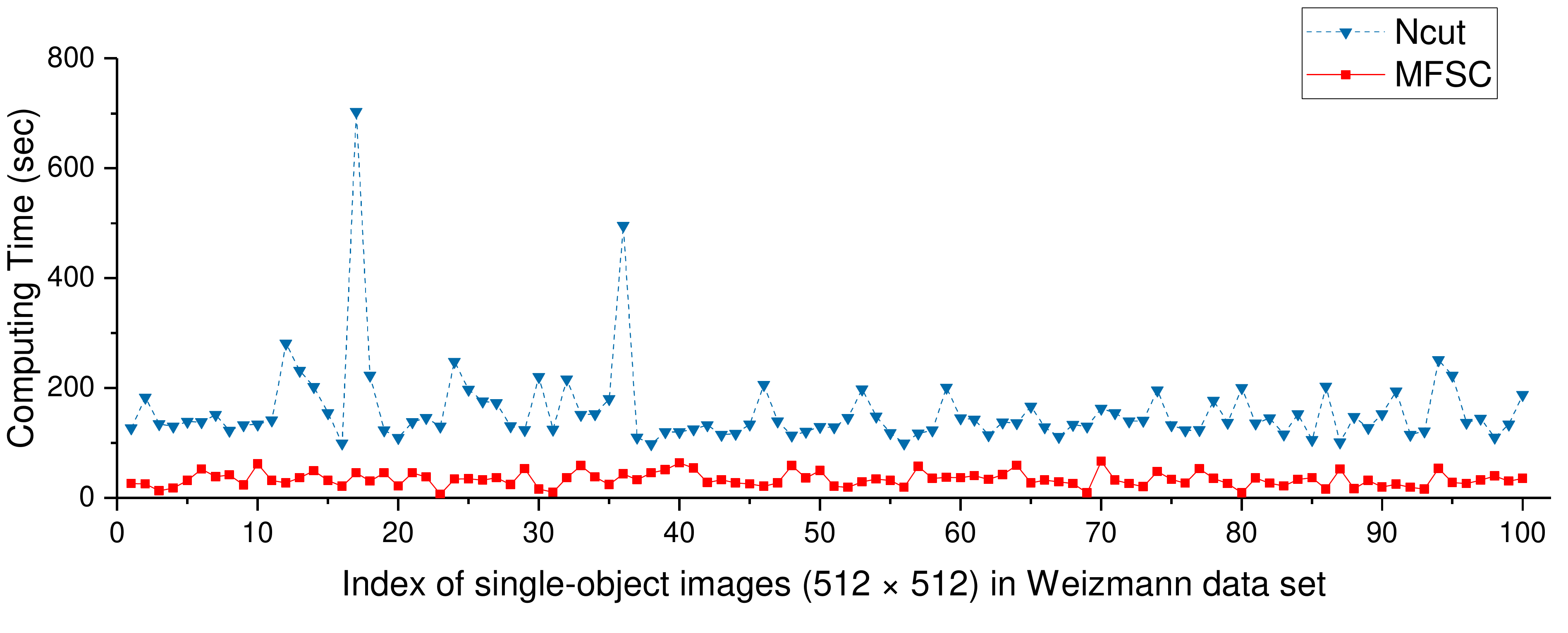}
        }
     \subfigure[]{
        \includegraphics[width=5in,height=2in]{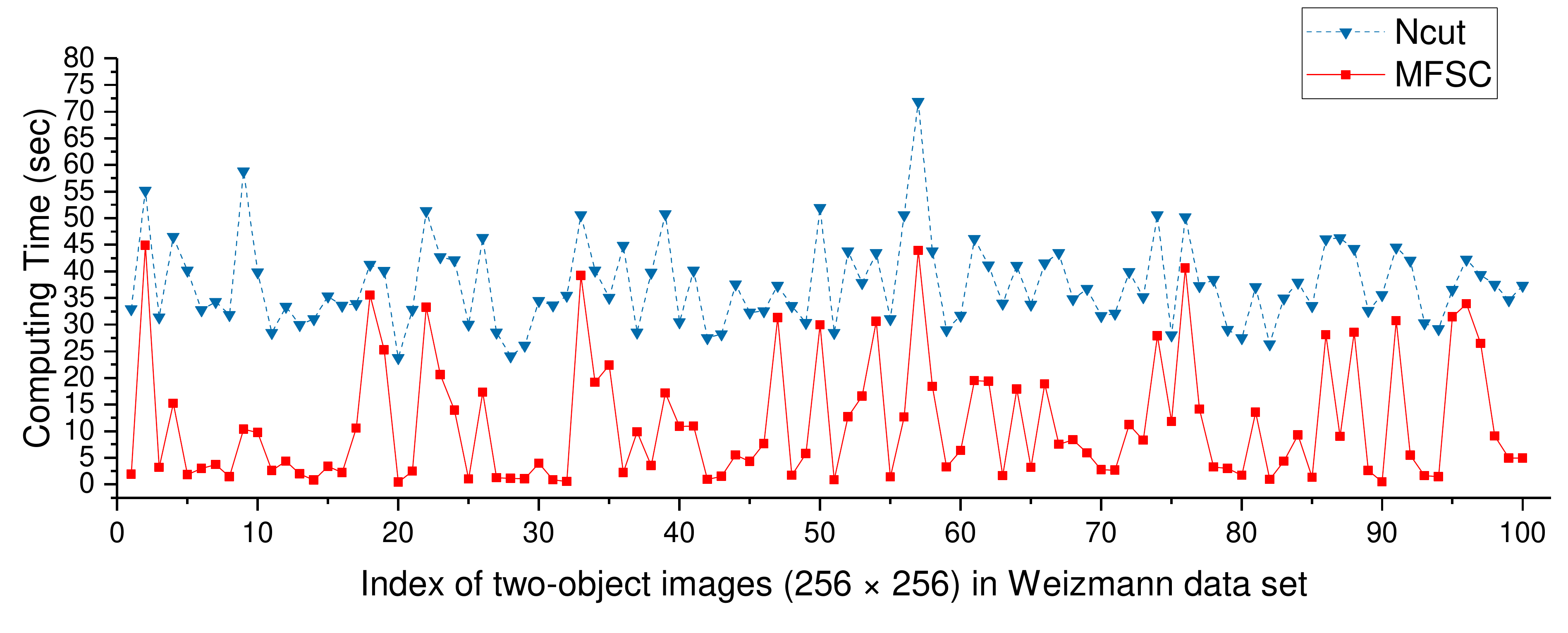}
        }
    \caption{The computing time of MFSC and Ncut on images in Weizmann data set. (a) single-object images ($128 \times 128$), (b) single-object images ($256 \times 256$), (c) single-object images ($512 \times 512$), (d) two-object images ($256 \times 256$).}
    \label{fig:three computing time total}
\end{figure*}

\subsection{Experimental Results of the Entire Weizmann Data Set}
\figurename \ref{fig:four image results} and \figurename \ref{fig:four Two object image results} show the segmentation results of the sample images in \figurename \ref{fig:four Image} and \figurename \ref{fig:object image} with MFSC and Ncut. Table \ref{table:performance in four images} and Table \ref{table:performance in four 2 obj images} present the segmentation accuracy of those images. We observe that the performances of MFSC and Ncut are similar in terms of accuracy. More importantly, by observing the computing time in Table \ref{table:running time in four image} and Table \ref{table:running time in four 2-obj image}, we know that MFSC outperforms Ncut in terms of efficiency.\par
\begin{table}[H]
\centering
\renewcommand{\arraystretch}{1.3}
\caption{Computing Time (Second) of the Four images displayed in Fig \ref{fig:four Image}}
\label{table:running time in four image}
\begin{tabular}{ccccc}
\hline
\bfseries Image name & \bfseries HotAirBalloon & \bfseries Nitpix &\bfseries Leafpav72 &\bfseries Tank\\
\hline\hline
\bfseries MFSC & 1.80 & 7.35 & 36.97 & 11.33\\
\bfseries Ncut & 7.28 & 29.87 & 138.93 & 115.88\\ \hline
\end{tabular}
\end{table}

\begin{table}[H]
\centering
\renewcommand{\arraystretch}{1.3}
\caption{Computing Time (Second) of the Four images displayed in Fig \ref{fig:object image}}
\label{table:running time in four 2-obj image}
\begin{tabular}{ccccc}
\hline
\bfseries Image name & \bfseries Plane & \bfseries Imgp1883 &\bfseries DualWindows &\bfseries Yack1\\
\hline\hline
\bfseries MFSC & 4.25 & 6.78 & 31.58 & 5.15\\
\bfseries Ncut & 32.29 & 41.52 & 41.98 & 34.63\\ \hline
\end{tabular}
\end{table}

To ensure that the comparison is fair, we use MFSC and Ncut on the Weizmann data set that contains 100 single-object images and 100 two-object images. The average computing time and the average accuracy of segmenting those images are reported in Table \ref{table:performance} and Table \ref{table:running time}. The graph radius and the threshold of the quad-tree that we use are as follows: For $128 \times 128$ single-object images, the parameters of MFSC are $R = 30$, $t = 10$, the parameter of Ncut is $r = 20$. For $256 \times 256$ single-object images, the parameters of MFSC are $R = 50$, $t = 12$, the parameter of Ncut is $r = 15$. For $256 \times 256$ two-object images, the parameters of MFSC are $R = 60$, $t = 8$, the parameter of Ncut is $r = 15$. For $512 \times 512$ single-object images, the parameters of MFSC are $R = 80$, $t = 15$, the parameter of Ncut is $r = 10$. As shown in Table \ref{table:performance}, for $512 \times 512$ images, the segmentation accuracy of MFSC is higher than that of Ncut. This is because on large-scale images, in order to obtain the segmentation result within the limited memory space and time, the graph radius of Ncut has to take a smaller value, with the result of sacrificing a certain amount of accuracy. \figurename \ref{fig:average time} shows the computing time of Ncut and MFSC on the images of different sizes. We observe that the computing time of Ncut increases steeply as the image becomes larger, whereas the time of MFSC rises gently. This experimental result is compliant with the computational complexity of MFSC that we've got in section \ref{subsection:Met-running time}. Finally, \figurename \ref{fig:three computing time total} shows the computing time of segmenting all of the 200 images of three sizes with MFSC and Ncut. As shown in \figurename \ref{fig:average time} and \figurename \ref{fig:three computing time total}, the computing time of MFSC is significantly shorter than that of Ncut.\par

%

%

%
%

\begin{table*}[!t]
\centering
\renewcommand{\arraystretch}{1.3}
\caption{The Average Segmentation Accuracy of MFSC and Ncut on Weizmann Data Set}
\label{table:performance}
\centering
\begin{tabular}{|c|c|c|c|c|c|c|c|c|c|c|c|c|}
\hline
\multirow{2}*{\diagbox{\bfseries Method}{\bfseries Data set}} & \multicolumn{3}{c|}{\bfseries single-object ($128 \times 128$)} &  \multicolumn{3}{c|}{\bfseries single-object in ($256 \times 256$)} &  \multicolumn{3}{c|}{\bfseries single-object ($512 \times 512$)} &  \multicolumn{3}{c|}{\bfseries two-object ($256 \times 256$)}\\
\cline{2-13}
                 & \bfseries ACC & \bfseries DICE & \bfseries RI & \bfseries ACC & \bfseries DICE & \bfseries RI & \bfseries ACC & \bfseries DICE & \bfseries RI & \bfseries ACC & \bfseries DICE & \bfseries RI\\
\hline
\bfseries MFSC & 0.85 & 0.72 & 0.78 & 0.83& 0.67 &0.74 &0.81 &0.64 &0.71 & 0.90 & 0.73 & 0.86\\

\hline
\bfseries Ncut & 0.83 & 0.71 & 0.76 & 0.81& 0.68 &0.74 &0.76 &0.58 &0.68 & 0.86& 0.73 & 0.81\\
\hline

\end{tabular}
\end{table*}
\begin{table*}[!t]
\centering
\renewcommand{\arraystretch}{1.3}
\caption{The Average Computing Time (Second) of MFSC and Ncut on Weizmann Data Set}
\label{table:running time}
\begin{tabular}{ccccc}
\hline
\bfseries Data set & \bfseries single-object ($128 \times 128$) & \bfseries single-object ($256 \times 256$) &\bfseries single-object ($512 \times 512$) &\bfseries two-object ($256 \times 256$) \\
\hline\hline
\bfseries MFSC & 2.25 & 14.09 & 33.28 & 11.28 \\
\bfseries Ncut & 8.14 & 36.65 & 157.36 & 37.46 \\ \hline
\end{tabular}
\end{table*}


\section{Conclusion and Future Work}
\label{section:conclusion}
In this paper, we present the Multiscale Fast Spectral Clustering algorithm for image segmentation. The results of our experiments on images of different sizes demonstrate the high efficiency of MFSC.\par
	In the future, we will expand the application of the method to color images and larger databases. We also plan to explore other methods to construct the hierarchical structure of the image.

\begin{IEEEbiography}[{\includegraphics[width=1in,height=1.25in,clip,keepaspectratio]{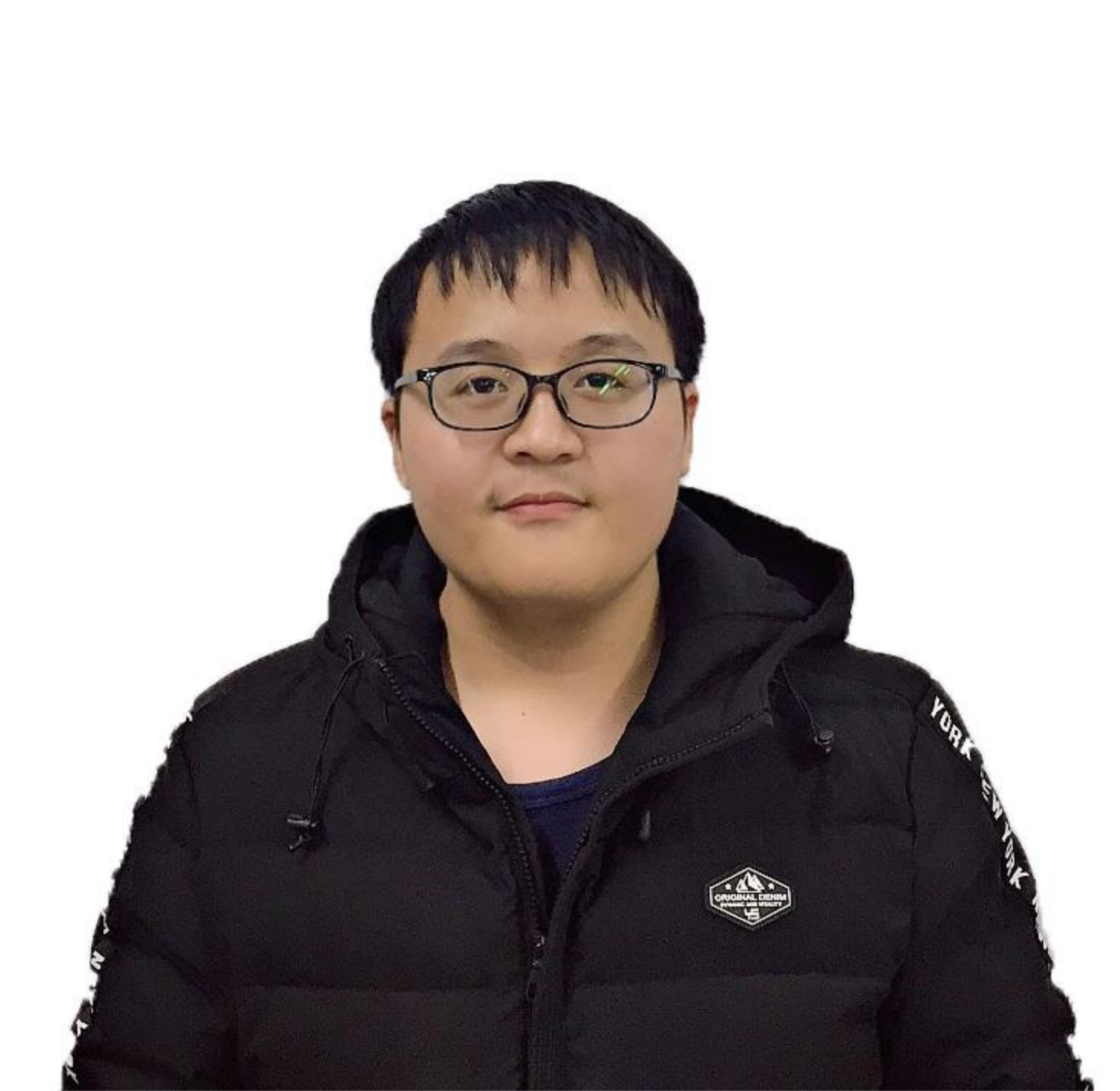}}]{Chongyang Zhang}
is currently pursuing the M.Eng. degree in the School of Electronic and Information Engineering from Soochow Univerity. His research interests is machine learning and image processing.
\end{IEEEbiography}

\begin{IEEEbiography}[{\includegraphics[width=1in,height=1.25in,clip,keepaspectratio]{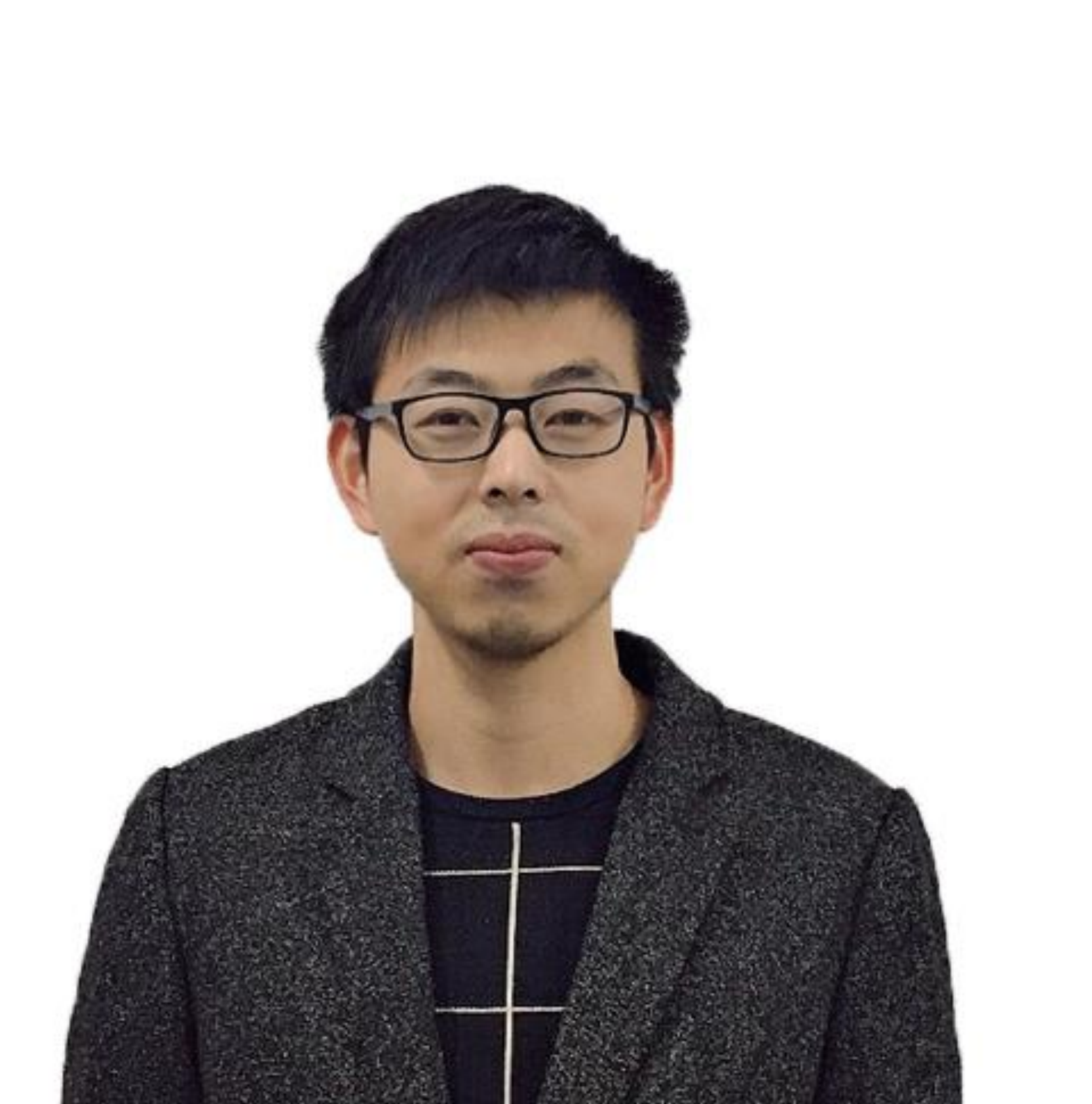}}]{Guofeng Zhu}
is currently pursuing the M.S. degree in the School of Mathematical Sciences from Soochow Univerity. His research interests is machine learning and image processing.
\end{IEEEbiography}

\begin{IEEEbiography}[{\includegraphics[width=1in,height=1.25in,clip,keepaspectratio]{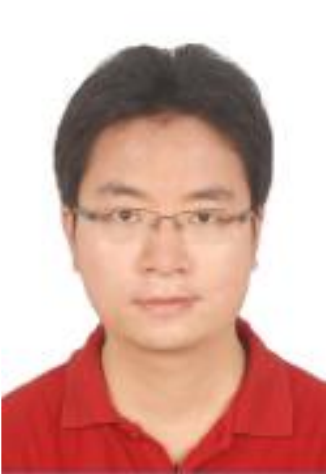}}]{Minxin Chen}
Minxin Chen is Associated Professor of computational mathematics at Soochow University, Suzhou, China. He received his Ph.D. in Computational mathematics at Chinese Academy of Sciences, Beijing, China. His research interests are in, image processing, and computational mathematics.
\end{IEEEbiography}

\begin{IEEEbiography}[{\includegraphics[width=1in,height=1.25in,clip,keepaspectratio]{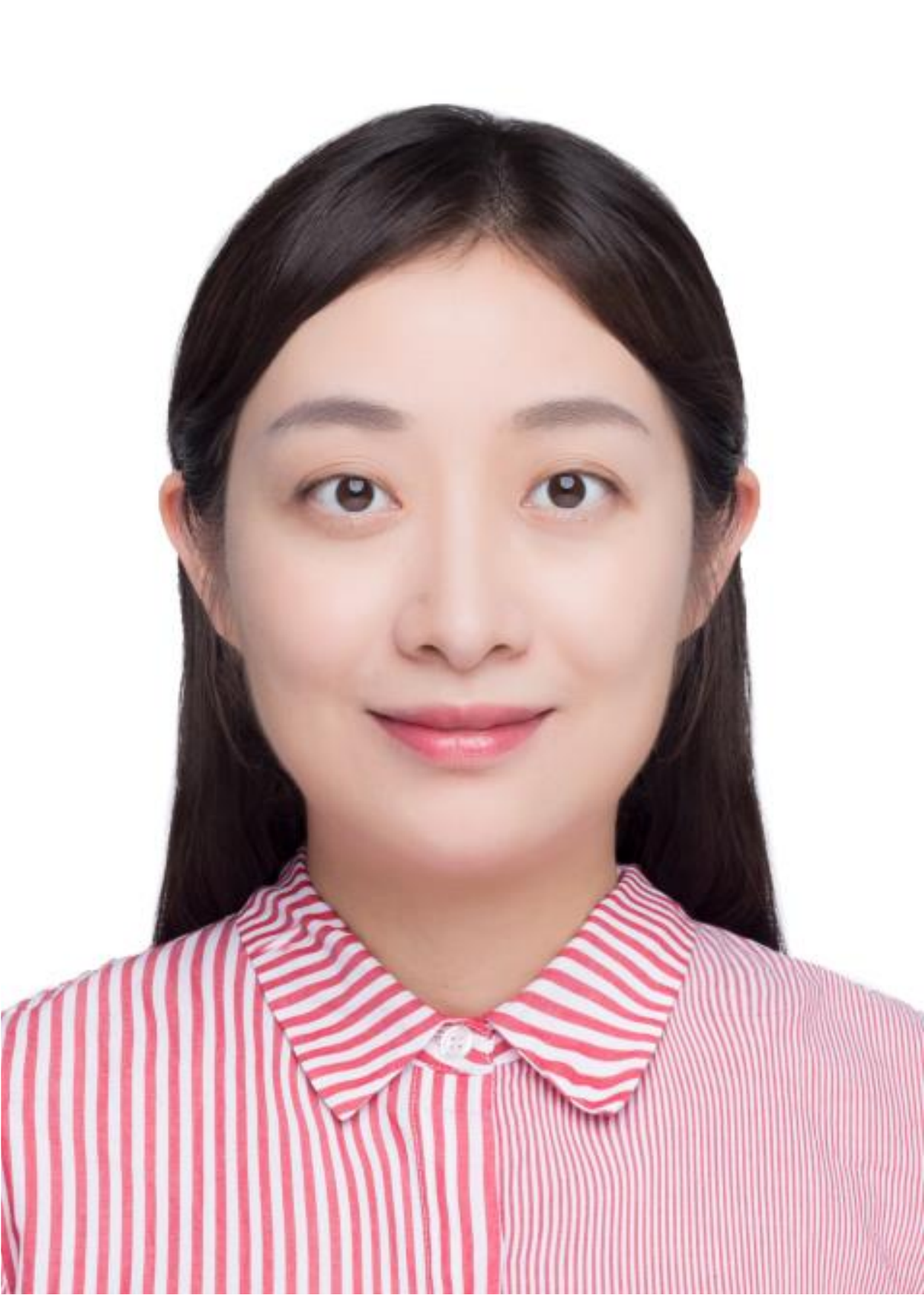}}]{Hong Chen}
Hong Chen was born in Suzhou, Jiangsu Province, China, in 1983.  She received her B.S.degree in School of Mathematical Sciences from Soochow University, Suzhou, China, in 2006, and she received her Ph.D degree in fundamental mathematics from Graduate University of Chinese Academy of Sciences, Beijing, China, in 2011. From 2011 until now, she worked in School of Mathematical Sciences from Soochow University.  She is currently working mainly on image processing and Mathematical modeling.
\end{IEEEbiography}

\begin{IEEEbiography}[{\includegraphics[width=1in,height=1.25in,clip,keepaspectratio]{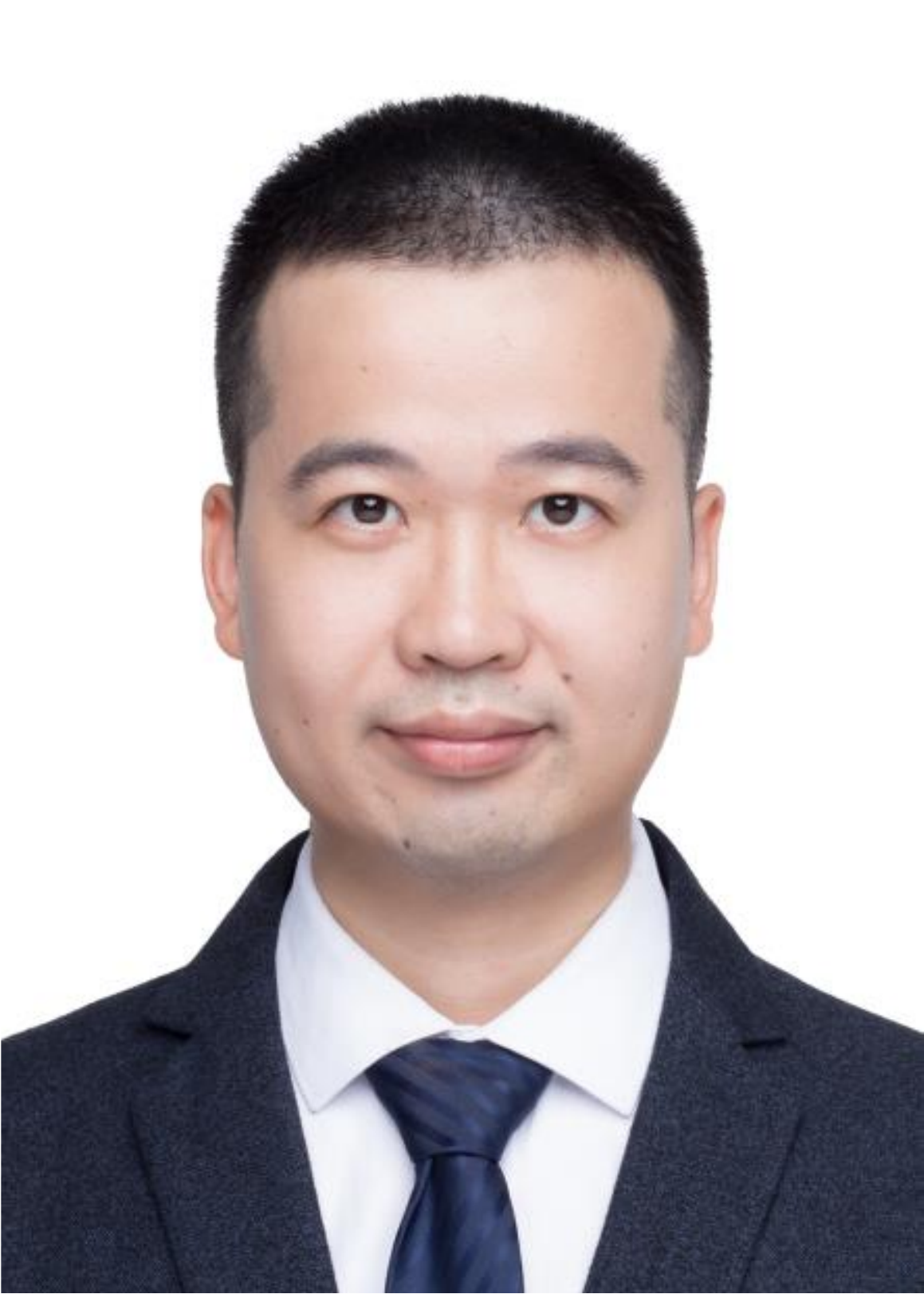}}]{Chenjian Wu}
Chenjian Wu was born in Suzhou, Jiangsu Province, China, in 1983.Chenjian Wu received his B.S. degree in Information Engineering from Southeast University, Nanjing, China, in 2006, and he received his M.S. degree in Software Engineering from Southeast University in 2010 and Ph.D degree in Electronic Circuit and System from Southeast University in 2013. From August 2013 until now, he worked in School of Electronic and Information Engineering, Soochow University, Suzhou, Jiangsu. He is currently working mainly on image processing and artificial intelligence chip design.
\end{IEEEbiography}

\end{document}